\documentclass[twocolumn,english,aps,prb,longbibliography,floatfix,prb,superscriptaddress]{revtex4-1}
\usepackage{amsmath,amssymb,natbib,bm,graphicx,url,epsfig,psfrag}
\usepackage{diagbox}
\usepackage{ulem}
\usepackage{makecell, multirow}
\usepackage[colorlinks=true,linkcolor=blue,urlcolor=blue,citecolor=blue]{hyperref} 
\hypersetup{colorlinks=true}
\usepackage[all]{hypcap} 

\usepackage{bbold}
\usepackage{color}

\begin{document}
\title{Symmetry-related transport on  a fractional  quantum Hall edge }

\author{Jinhong Park}
\affiliation{Department of Condensed Matter Physics, Weizmann Institute of Science, Rehovot 76100, Israel}
\affiliation{
Institute for Theoretical Physics, University of Cologne, Z{\"u}lpicher Str. 77, 50937 K{\"o}ln, Germany}
\author{Bernd Rosenow}
\affiliation{Institut f{\"u}r Theoretische Physik, Universit{\"a}t Leipzig, D-04103 Leipzig, Germany}
\affiliation{Department of Condensed Matter Physics, Weizmann Institute of Science, Rehovot 76100, Israel}
\author{Yuval Gefen} 
\affiliation{Department of Condensed Matter Physics, Weizmann Institute of Science, Rehovot 76100, Israel}
\date{\today}

\begin{abstract}

 Low-energy transport in quantum Hall states is carried through edge modes, and is dictated by bulk topological invariants and possibly microscopic Boltzmann kinetics at the edge. Here we show how the presence or breaking of symmetries of the edge Hamiltonian underlie transport properties, specifically d.c.~conductance and noise. We demonstrate this through the analysis of hole-conjugate states of the quantum Hall effect, specifically the $\nu=2/3$ case in a quantum point-contact (QPC) geometry. We identify two symmetries, a continuous $SU(3)$ and a discrete $Z_3$, whose presence or absence (different symmetry scenarios) dictate qualitatively different types of  behavior of conductance and shot noise.  While recent measurements are consistent with one of these symmetry scenarios, others can be realized in future experiments.

\end{abstract}

\maketitle

\section{Introduction}

\subsection{Questions we address}

The edge of  quantum Hall (QH) phases supports gapless excitations. These are responsible for low energy dynamics in such systems, including electrical and thermal transport and noise. A convenient working framework to study this physics is to describe the edge in terms of one-dimensional chiral Luttinger modes~\cite{WEN1992}. This simple picture has proven more complex and exotic than first anticipated, especially in the context of fractional quantum Hall (FQH) phases~\cite{Bid2010,Gurman2012, Gross2012,Venkatachalam2012, Altimiras2012, Inoue2014, Takei2011, Viola2012, Shtanko2014,Bid2009,Takei+2015, Sabo2017, Rosenblatt2017,Das2019}.
This is a particularly pressing issue when it comes to multi-mode edges, e.g., the case of hole-conjugate states ($1/2 < \nu< 1$), where counter-propagating chiral modes are present. Varying the confining potential at the edge, and accounting for the effective electrostatic and exchange interaction may give rise to edge reconstruction~\cite{Chklovskii1992, Dempsey1993, Chamon1994}, where additional chiral edge modes emerge~\cite{Meir1994, Wang2013, MacDonald1990, Sondhi1993, Yang2003, Wan2003, Wan2002, Khanna2020}. Topological numbers of the bulk phase impose constraints on these emergent modes, importantly that the difference between the number of upstream and downstream chirals (proportional to the topological heat conductance~\cite{Kane1997}) remains invariant under edge reconstruction. 

The interplay of disorder and interactions at the edge may lead to renormalization of the edge modes. One paradigmatic example is the edge of a bulk filling factor $\nu_{\rm bulk}=2/3$, where a downstream charge 2/3 modes and an upstream neutral mode may emerge~\cite{Kane1994, Kane1995,Rosenow2010,Protopopov2017,Nosiglia+2018}. At the stable fixed point the charge and neutral sectors are decoupled; the latter is underlied by an $SU(2)$ symmetry, related to the fact that neutral excitations (“neutralons”) may be mapped onto spin-1/2 fermions.

The Kane-Fisher-Polchinski model~\cite{Kane1994} serves as an example of an emergent symmetry ($SU(2)$) due to renormalization group (RG) of edge modes. This evokes a much broader scoped question: the interplay of topology and emergent symmetry, and its role in dictating transport and noise at the edge. To address this question we focus here on a paradigmatic example, that of a reconstructed and renormalized edge of the $\nu=2/3$ FQH phase.  Indeed, the predicted upstream neutral mode propagation~\cite{Kane1994} has been observed experimentally~\cite{Gurman2012, Venkatachalam2012, Altimiras2012, Inoue2014,Bid2010,Gross2012, Cohen2019}. Subsequent experiments~\cite{Bid2009, Sabo2017, Rosenblatt2017, Bhattacharyya2019} implementing quantum point contacts (QPCs), led to the conclusion that the original picture of the edge~\cite{MacDonald1990, Meir1994, Kane1994} needed to be modified~\cite{Wang2013}. Following a RG analysis of the reconstructed edge (Fig.~\ref{edgest1}(a)), the emerging picture comprises an intermediate fixed point with two downstream running charged modes and two upstream neutral modes (Fig.~\ref{edgest1}(b)). While much of the available experimental data is compatible with this edge structure (e.g., the conductance plateau of $e^2 / 3h $ and the existence of upstream modes), the observation of significant shot noise (corresponding to the Fano factor $2e/3$) on the conductance plateau~\cite{Bid2009, Sabo2017, Bhattacharyya2019} is still puzzling. The conductance plateau implies that each of the charge modes either fully reflected or fully trasmitted. Such a scenario would suggest the absence of shot noise on the plateau.~\cite{footnote1}

Moreover the robustness of (and evidence for) incoherent transport~\cite{Rosenow2010, Takei2011,Takei+2015, Protopopov2017, Nosiglia+2018, Park2019, Spanslatt2019, Spanslatt2020}, described by microscopic Boltzmann kinetics of edge modes, underlies most experimental studies. Further experimental evidence of coherent upstream propagation~\cite{Protopopov2017} of neutral  modes (underlied by intriguing hidden symmetries as discussed below) is needed.

It is clearly desirable to incorporate our current understanding of edge structure (dictated by topological constraints) and edge dynamics within a  general symmetry based framework. This is the main goal of the  present work. While we focus here on the $\nu=2/3$ reconstructed and renormalized edge, the ideas presented here can serve as guidelines in studying other topological states. Furthermore, our study may motivate the search for various symmetry scenarios with new experimental manifestations.

\subsection{Summary of main results}

\begin{figure} 
\includegraphics[width=\columnwidth]{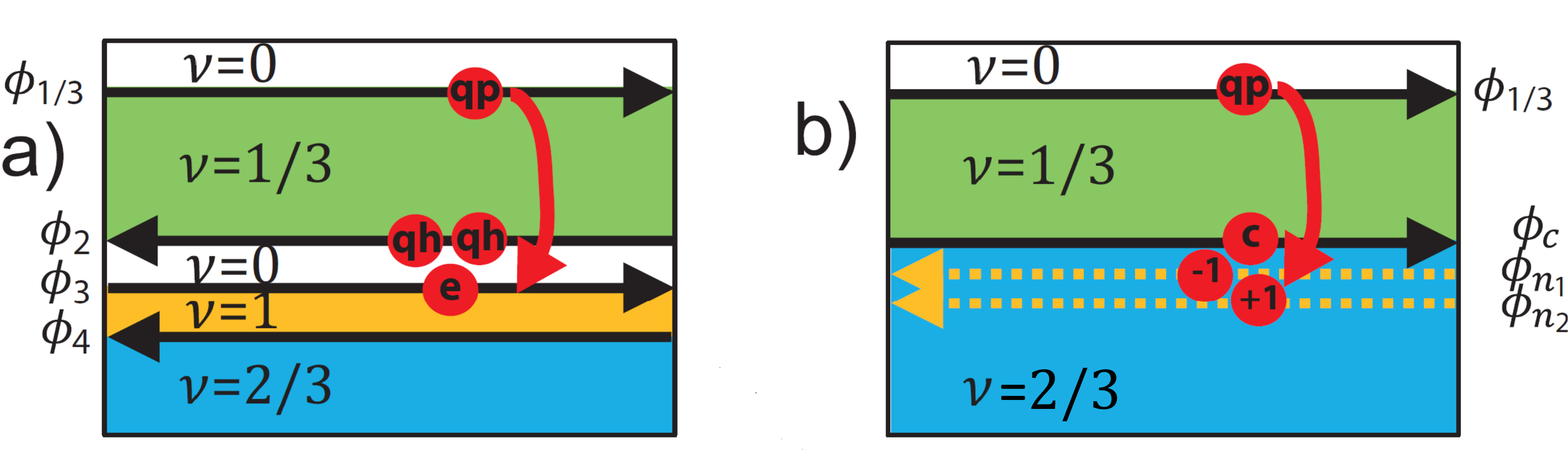} 
\caption{(color online). Edge structures and excitations of the bulk filling factor $\nu_{\rm{bulk}} = 2/3$.  a) 
Four bare edge modes are depicted with arrowheads indicating the direction
of each mode (downstream (right movers) or upstream (left movers)). A quasi-particle  tunneling from the outermost edge mode to the inner ones turns into two quasi-holes and one electron.  Due to the presence of a vacuum region with $\nu=0$, only electrons can tunnel to mode $\phi_4$, leaving the fractional charge on $\phi_4$ invariant, and giving rise to a $Z_3$ symmetry between
sectors with fractional charge $0$, $1/3$, and $2/3$. 
b) Due to electron tunneling between the three inner edge modes ($\phi_2$, $\phi_3$ and $\phi_4$) and bare interaction between them, 
the inner three edge modes are renormalized to
a downstream charge mode (denoted as a black solid line) with the filling factor discontinuity $\delta \nu = 1/3$ and 
two upstream neutral modes (denoted as yellow dashed lines). The same tunnel process as in a) now creates a chargeon and neutral charges $(-1,1)$.  
}\label{edgest1}
\end{figure}

\begin{figure} 
\includegraphics[width=\columnwidth]{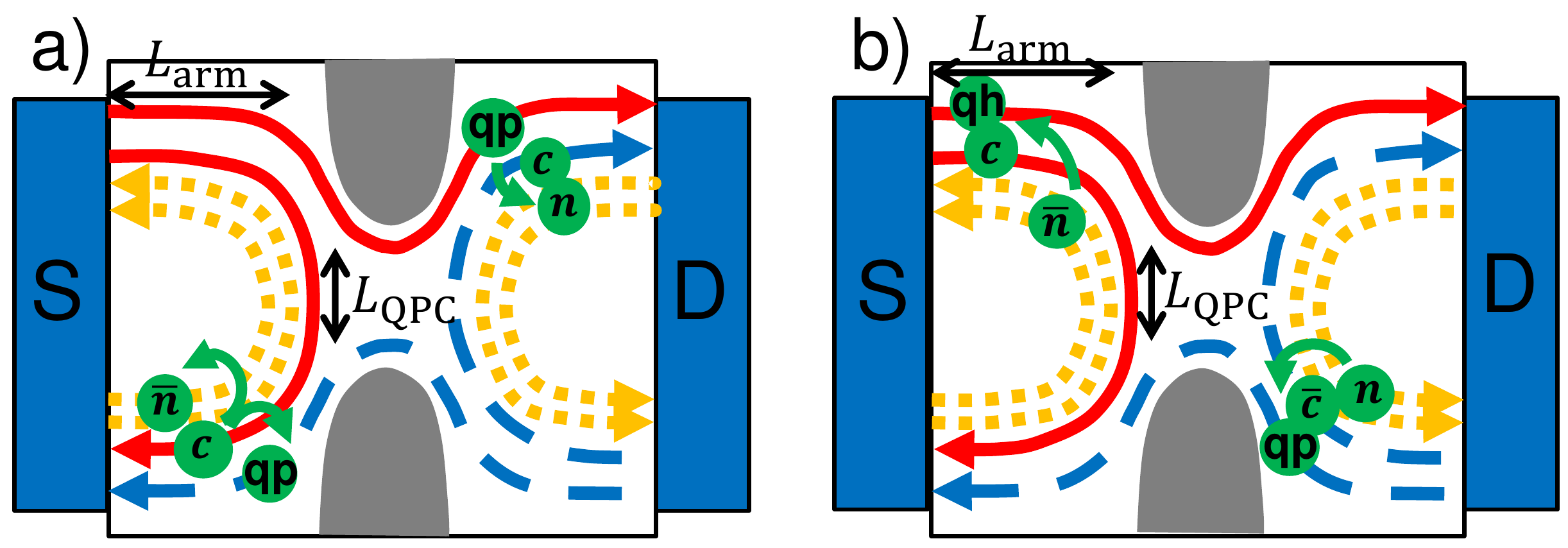} 
\caption{ (color online). A sketch of a two-step mechanism contributing to the conductance in a two terminal setup with a quantum point contact (QPC). 
  Potentials are tuned such that the outermost mode is fully transmitted through the QPC while the inner ones are fully reflected. The outer charge mode supports quasi-particles (qp) and quasi-holes (qh), while the inner one supports chargeons ($c$) and anti-chargeons ($\bar{c}$), all four carrying a charge of $\pm e/3$. Similarly, the two neutral modes support neutral excitations, neutralons ($n$) or anti-neutralons ($\bar{n}$). The size of the QPC region and the length of the arms between the contacts and the QPC are $L_{\rm{QPC}}$ and $L_{\rm{arm}}$, respectively.
(a) Equilibration processes involving charge tunneling and the creation of neutralons or anti-neutralons: $\textrm{qp} \rightarrow c+ n$  (upper right corner) or $c \rightarrow \textrm{qp}+ \bar{n}$ (lower left corner) represented in Eq.~\eqref{qptunnelingeq}, with $
\ell_{\rm ch-eq, 0}\ll L_{\rm{arm}}$.
(b) The excited (anti-)neutralons arrive at the lower right (upper left) corner and they decay giving rise to qp/$\bar{c}$ (qh/c)  pairs. The excited qp, qh, c, or $\bar{c}$ flow to the drain D, contributing to the conductance. In the presence of  
$H_{n \bar{n}}$ Eq.~\eqref{mixing}, mixing between neutralons and anti-neutralons occurs over a distance $\ell_{n \bar{n}, 0} \ll L_{\rm QPC}$.  
}\label{Process2}
\end{figure}

The present work demonstrates that, given the topological invariants  of the bulk phase  (dictated by the filling factor),  different symmetries of the edge modes may  underlie qualitatively different transport behavior, specifically the d.c. conductance and the low-frequency non-equilibrium noise. This facilitates the engineering of experimentally controlled setups (of the same bulk phase) with designed, symmetry-related, behavior. In order to demonstrate our approach, we consider a specific geometry, write down the most general fixed point action compatible with this setup, identify the relevant symmetries, and show how their presence (or absence) determine the resulting transport properties.

We start by identifing two symmetries of the $\nu=2/3$ edge, whose presence or absence (different symmetry scenarios) dictate qualitatively different types of behavior of the conductance and the shot noise: a continuous $SU(3)$ symmetry and a discrete $Z_3$. The former symmetry acts in the subspace spanned by the neutral modes. Inspired by high energy terminology, we refer to the basic neutral excitations as {\it up}, {\it down} and {\it strange} (cf. Eq.~\eqref{Neutralsector2} and the discussion thereafter). The space of these {\it neutralons} forms the fundamental representation of an $SU(3)$ symmetry group. The conjugate representation is associated with {\it anti-neutralons}.  At the fixed point, the  action is invariant under $SU(3)$ operations; as discussed below, one may break this symmetry. If this symmetry breaking occurs in spatially randomly varying and self-averaging, one can refer to a "statistically preserved" $SU(3)$ symmetry. In physical terms, the breaking of the $SU(3)$ symmetry takes place through charge equilibration in the charge sector of the chiral modes (see Fig.~\ref{edgest1}(b)). The second important group here is a discrete $Z_3$ (cf. Eq.~\eqref{mixing}, related to three different sectors of the charge on $\phi_4$  with the fractional charge $0$, $1/3$ or $2/3$. This group is not to be confused with the $Z_3$ subgroup of the $SU(3)$.
Breaking this  $Z_3$ symmetry allows to connect the neutralon sector with the anti-neutralon sector. This may be achieved by the tunneling of $1/3$ charges into the innermost bare edge mode (cf. Fig.~\ref{edgest1}(a)).

To see how the symmetries affect transport behavior in a specific configuration, we consider a two terminal setup (cf. Fig.~\ref{Process2}) with a QPC. The edge of the system consists of the edge modes depicted in Fig.~\ref{edgest1}(b). The QPC is  tuned such that the outermost charge mode is fully transmitted  while the inner ones are fully reflected, naively leading to a
 quantized conductance of $(1/3)(e^2/h)$. Through charge equilibration between the charge modes (cf.~Fig.~\ref{Process2}(a) upper right) that breaks the $SU(3)$ symmetry of the fixed point, only one type of neutral excitations (neutralons, but not anti-neutralons)  is created. These neutralons then  decay converting  into quasi-particles in the outermost mode and quasi-holes in the inner charge mode (cf.~Fig.~\ref{Process2}(b) lower right), generating an additional current at drain D, 
 hence undermining conductance quantization (cf.~the upper row in Table I). 
%
Breaking the $Z_3$ symmetry gives rise to the emergence of a quantized conductance, $(1/3)(e^2/h)$. 
The reason is that with the $Z_3$ symmetry broken, the generation of neutralons and anti-neutralons in the course of equilibration is equally probable, implying no additional d.c. current at drain $D$ yet a contribution to shot noise with a quantized Fano factor, $2 e/3$. Experimental manifestations of the various symmetry scenarios are summarized in Table~\ref{table1}.

\subsection{Structure of the paper}

The outline of this paper is the following. In Section II we describe a $SU(3)$ fixed point emerging at intermediate energies. In Section III, we consider tunneling operators near the fixed point, each of which break or preserve the $SU(3)$ or $Z_3$ symmetry, and elaborate specific models for the two terminal setup.
In Section IV, we analyze transport properties (d.c. conductance and non-equilibrium noise), demonstrating how they are affected by presence or absence of centain symmetries. Section V is a summary.

\section{$SU(3)$-symmetric fixed point} 

To make this discussion more self-contained, in the following we give an overview over the derviation of the $SU(3)$ symmetric intermediate fixed point action \cite{Wang2013} describing the inner three edge modes. We start by considering the filling fraction profile sketched in Fig.~1(a), consisting of a $\nu=1/3$ incompressible region near the boundary of the sample, separated by more narrow  $\nu=0$ and $\nu=1$ regions from the bulk. The drop of the filling fraction 
to zero is dictated by topology, in the sense that the $1/3$ FQH state is topologically distinct from the $2/3$ state, and hence a direct transition between them is not possible. Such a filling fraction profile is compatible with all topolgical constraints and can in principle be stabilized by a suitably chosen external potential. 
Modifications making this profile more realistic are described via the inclusion of interaction and scattering terms.

Edge states arise at the boundary of two incompressible regions, such that there is an outermost  downstream $1/3$ edge mode described by a boson field $\phi_{1/3}$, rather well separated from an upstream $1/3$ mode described by a field $\phi_2$, a downstream integer mode described by $\phi_3$, and another upstream $1/3$ mode described by $\phi_4$.  We want to focus on the inner three edge modes at the moment, and assume that they are spatially in close proximity, much closer to each other than to the outermost 1/3 mode.  Due to the presence of the integer incompressible region between the two fractional modes, only electron scattering between them is possible. Although electron scattering is irrelevant in the absence of Coulomb interaction, spatially random scattering becomes relevant in the presence of sufficiently strong interactions, and the RG flows  towards strong coupling for the electron tunneling~\cite{Wang2013}. 

In order to describe the disorder dominated strong coupling fixed point, one next makes an educated guess for a fixed point action, consisting of a charge mode $\phi_c \equiv (\phi_2 + \phi_3 + \phi_4) /\sqrt{3} $ and two neutral modes $\phi_{n_1}\equiv  (3\phi_2 + \phi_3)/\sqrt{2}$ and $\phi_{n_2}\equiv(\phi_2 + \phi_3 + 2\phi_4)/\sqrt{2}$:
\begin{widetext}
 \begin{align} \label{Neutralsector}
 S_0 &= S_c + S_n, \nonumber \\ 
 S_c & = \frac{1}{4\pi }\int dx dt  \left (\partial_x \phi_{c} (-\partial_{t}-v_{c}\partial_x ) \phi_{c} 
 \right ) \nonumber \\ 
 S_{n} &=\int dx dt \bigg [ \frac{1}{4\pi} \left (\partial_x \phi_{n_1} (\partial_{t}-v_{n}\partial_x ) \phi_{n_1}
+3\partial_x \phi_{n_2} (\partial_{t}-v_{n}\partial_x ) \phi_{n_2} \right ) \nonumber \\ &
-\frac{1}{2\pi a}\left (\xi_{1}(x) e^{i \sqrt{2} \phi_{n_1}}+ \xi_{2}(x) e^{- i \phi_{n_1} /\sqrt{2}}  e^{3 i \phi_{n_2}/\sqrt{2}} 
+\xi_{3}(x)  e^{ i \phi_{n_1}/\sqrt{2}}  e^{3 i \phi_{n_2} /\sqrt{2}}  + \textrm{H.c.} \right ) \bigg ].
 \end{align}
 \end{widetext}
The charge and neutral modes satisfy commutation relations
$[\phi_{c}(x), \phi_{c}(x')] = i\pi \textrm{sgn}(x-x')$,
$[\phi_{n_1}(x), \phi_{n_1}(x')] = -i\pi \textrm{sgn}(x-x')$, and
$[\phi_{n_2}(x), \phi_{n_2}(x')] = -i\pi \textrm{sgn}(x-x')/3$.
The second line originates from  electron tunneling between the inner modes. $\xi_{i = 1, 2, 3}$ are random variables. 
We note that the six electron tunneling operators in the second line of the above equation, together with the operators $\partial_x \phi_{n_1}$ and 
$\partial_x \phi_{n_2}$ constitute the generators of the symmetry group $SU(3)$, in the sense that their commutation relations form the $SU(3)$ 
algebra. One can show that the action Eq.~\eqref{Neutralsector} indeed describes an attractive fixed point in the sense that all interactions terms between charge mode and neutral modes, and among neutral modes, flow to zero. 

In order to make the $SU(3)$ symmetry more explicit, we next fermionize the action Eq.~\eqref{Neutralsector}. 
The idea is to introduce an auxiliary boson which does not couple to the 
two neutral mode bosons $\phi_{n_1}$, $\phi_{n_2}$, and use the two neutral modes and the auxiliary boson to fermionize the model. 
Including the action for an auxiliary field $\chi$ (see Ref.~\cite{Kane1994} for a similar procedure)
and performing fermionization in terms of a three-component fermion field $\Psi (x) =  e^{i \chi /\sqrt{3}}  [e^{-i(\phi_{n_1}+\phi_{n_2})/\sqrt{2}}, e^{i(\phi_{n_1}-\phi_{n_2})/\sqrt{2}}, e^{i\sqrt{2}\phi_{n_2}}]^T /\sqrt{2 \pi a} = e^{i \chi / \sqrt{3}} [\psi_{u}, \psi_{d}, \psi_{s}]^T$, Eq.~\eqref{Neutralsector} reads
%
 \begin{widetext}
  \begin{align} \label{Neutralsector1}
 S_{n} &= \int dx dt \left [ i\Psi^{\dagger} (-\partial_t +v_n \partial_x) \Psi - \left ( \xi_{1} \Psi^{\dagger} (\frac{\lambda_1 + i \lambda_2}{2}) \Psi + \xi_{2} \Psi^{\dagger} (\frac{\lambda_6 + i \lambda_7}{2}) \Psi+ \xi_3 \Psi^{\dagger} (\frac{\lambda_4 + i \lambda_5}{2})\Psi + \textrm{H.c.} \right ) \right].
 \end{align}
 \end{widetext}
 %
 Here, the matrices $\lambda_{1,2,4,5,6,7}$ are non-diagonal generators of the $SU(3)$ group 
(see Appendix~\ref{appen_SU3} for more details).
 
 \begin{figure} 
\includegraphics[width=.8\columnwidth]{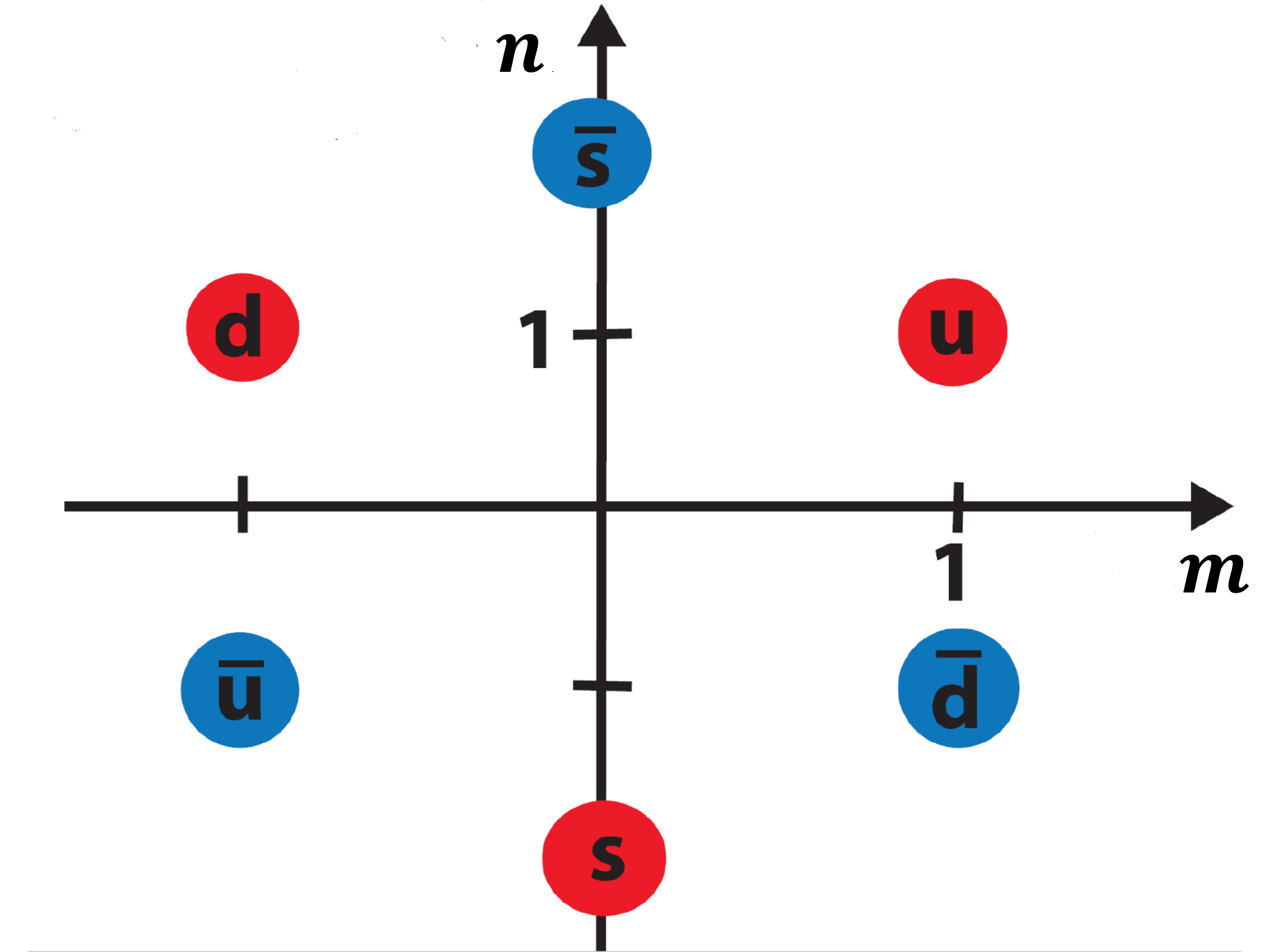} 
\caption{(color online). $SU(3)$ group representations. Quasi-particles described by charges $(m, n)$ (cf. Eq.~\eqref{charges}) form the fundamental representation of $SU(3)$. In analogy to the quantum numbers $(I_z, Y)$ of isospin and hypercharge in flavor $SU(3)$ \cite{su3book}, the elementary excitations (neutralons) are 
labeled $u$, $d$, and $s$.  The corresponding anti-neutralons form the conjugate representation; a conversion of  neutralons into anti-neutralons is only possible if the $Z_3$ symmetry is broken. This can be achieved by allowing fractional quasiparticles to tunnel into and out of the inner-most edge mode. More details pertaining to the $Z_3$ symmetry can be found in Eq.~\eqref{correlator} and the discussion thereafter, and the $SU(3)$ group lattice is described in Appendix~\ref{appen_SU3}.
}\label{su3lattice}
\end{figure}
 
 In the fermion picture, the tunneling corresponds to random rotations between the fermions, and can be transformed away by choosing a new basis via a space-dependent $SU(3)$ rotation.
 The random terms are completely eliminated via performing a $SU(3)$ gauge transformation of 
 $\tilde{\Psi} (x) = U(x) \Psi (x) $; the action becomes diagonal as 
 \begin{align}
 S_{n} = i \int dx dt \tilde{\Psi}^{\dagger} (-\partial_t + v_n \partial_x ) \tilde{\Psi}.
 \end{align}
 Here, 
 %
  \begin{equation} \label{rotationmatrix}
  U(x) = T_x  e^{-{ i\over 2 v_n} \int_{-\infty}^{x} dx' \left [\xi_{1}\left (\lambda_1 + i \lambda_2 \right) + \xi_2 \left( \lambda_6 + i \lambda_7 \right ) +
 \xi_3 \left (\lambda_4 + i \lambda_5 \right)  + \textrm{H.c.}\right ]} 
 \end{equation}
 %
denotes the position dependent $SU(3)$ rotation matrix. Transforming away the disorder in this way makes the velocities all equal. 
 
 The introduction of  the auxiliary boson has contributed an additional $U(1)$ symmetry to the problem. This extra symmetry can be removed by performing a 
 rebosonization of the rotated fermionic fields as  $\tilde{\Psi}^{T}= e^{i \tilde{\chi} /\sqrt{3}} [\tilde{\psi}_{u}, \tilde{\psi}_{d}, \tilde{\psi}_{s}]$. Here 
 \begin{align}
 \tilde{\psi}_{u} &\equiv \frac{1}{\sqrt{2 \pi a}} e^{-i(\tilde{\phi}_{n_1}+\tilde{\phi}_{n_2})} ,\nonumber \\ 
\tilde{\psi}_d &\equiv  \frac{1}{\sqrt{2 \pi a}}e^{i(\tilde{\phi}_{n_1}-\tilde{\phi}_{n_2})},\nonumber \\ 
\tilde{\psi}_s  &\equiv  \frac{1}{\sqrt{2 \pi a}}e^{i\sqrt{2}\tilde{\phi}_{n_2}}
\end{align}
are referred to as {\it up}, {\it down}, and {\it strange} neutralon in the present paper.
In this way, and omitting the tilde, Eq.~\eqref{Neutralsector1} becomes
 %
 \begin{eqnarray} 
 S_{n} &= &\frac{1}{4\pi} \int dx dt  \left[ (\partial_x \phi_{n_1} (\partial_{t}-v_{n}\partial_x ) \phi_{n_1} \right. \nonumber \\
& &\left. +3\partial_x \phi_{n_2} (\partial_{t}-v_{n}\partial_x ) \phi_{n_2}  ) \right]   \ .  \label{Neutralsector2}
 \end{eqnarray}
 %
The corresponding Hamiltonian together with the Hamiltonian related to the charge modes is given by 
\begin{align}  \label{bareHamilonian}
H_0 = H_n + H_c  =& \frac{v_n}{4 \pi}\int dx \left [ (\partial_x \phi_{n_1})^2 +3(\partial_x \phi_{n_2})^2 
\right ] \nonumber \\ +&\frac{1}{4 \pi}\int dx \left [v_c (\partial_x \phi_{c})^2 +v_{1/3}(\partial_x \phi_{1/3})^2 
\right ].
\end{align}
Two representations of the $SU(3)$ group are displayed in Fig.~\ref{su3lattice}. 
 In general, neutralon states, which form the fundamental representation of the $SU(3)$ group, are labeled by charges $(m, n)$
that are eigenvalues of the diagonal generators $\lambda_3$ and $\lambda_8 \sqrt{3}$, respectively; 
\begin{align} \label{charges}
\lambda_3 |j \rangle=m |j \rangle, \,\,\,\,\,\,\,\, \lambda_8 \sqrt{3} |j\rangle= n |j \rangle.
\end{align} with 
 \begin{align} \label{diagonalgenerator}
\lambda_3 = \begin{pmatrix}
      1 & 0 & 0\\
   0 & -1 & 0 \\
  0  & 0 & 0  
 \end{pmatrix},\,\,\,\,\,\,\, 
\lambda_8 = \frac{1}{\sqrt{3}}\begin{pmatrix}
      1 & 0 & 0\\
   0 & 1 & 0 \\
  0  & 0 & -2  
 \end{pmatrix}.
\end{align}
The neutralon states of Fig.~\ref{su3lattice}  (denoted as red dots) are denoted as 
$|u \rangle = (1, 0, 0)^T = \psi_u^{\dagger} |0 \rangle$, $|d \rangle= (0, 1, 0)^T= \psi_d^{\dagger} |0 \rangle$ and $|s \rangle= (0, 0, 1)^T=\psi_s^{\dagger} |0 \rangle$, which 
correspond to $(m, n) = (1, 1), (-1,1)$ and $(0,-2)$ respectively. 
Likewise, the conjugate representation is associated with 
 the anti-neutralons ($\bar{u}$, $\bar{d}$, and $\bar{s}$) (depicted as blue dots in Fig.~\ref{su3lattice}). At the intermediate $SU(3)$ fixed point, the neutralon sector is completely disconnected from the anti-neutralon sector. In the next section, we will show that breaking a $Z_3$ symmetry makes a connection between the sectors. More details for the $SU(3)$ group structure can be found in Appendix~\ref{appen_SU3}.


%

\section{Tunneling operators and models}

At the intermediate fixed point,  there is neither charge nor thermal equilibration between the different modes, implying 
a non-zero thermal conductance. We now want to describe equilibration between the charge modes (which also includes  the creation of neutral excitations), but still assume that the size of the system is shorter than the inelastic scattering length, at which the edge thermal conductance decays towards zero (see Appendix~\ref{app_length} for a justification of this assumption by a RG analysis). To this end,  we consider quasi-particle tunneling ("charge equilibration") between the outer-most mode and the three inner modes, expressed in terms of the most relavant operators:
\begin{align} \label{qptunnelingeq}
H_{\rm{ch-eq}}= & 
  \int dx \Big [ (t_{u} \psi_{u}^{\dagger} + t_{d}  \psi_{d}^{\dagger}
+ t_{s}  \psi_{s}^{\dagger} ) e^{- i \phi_c /\sqrt{3}} e^{ i\phi_{1/3}}  
\nonumber \\ &
+ \textrm{H.c.} \Big].
\end{align}
Note that $H_{\rm{ch-eq}}$ breaks the $SU(3)$ symmetry.  A quasi-particle $e^{i \phi_{1/3}}$ is annihilated, creating a {\it chargeon}  $e^{- i \phi_c /\sqrt{3}}$ (the charge  sector of a quasi-particle excitation carrying a charge $e/3$ formed in the inner modes) and a neutralon. Conversely, while a quasi-particle is created, a chargeon is annihilated together with the creation of an {\it anti-neutralon} (the anti-particle of the neutralon).
The microscopic origin of Eq.~\eqref{qptunnelingeq} is presented in Appendix~\ref{appen_tunnelingop}.

We next recover the $SU(3)$ symmetry through disorder averaging. The tunneling amplitudes $t_{j = u, d, s}$ are random with white noise correlation function $\langle t_{j_1} (x) t_{j_2}^{*} (x') \rangle_{\textrm{dis}} = W_{\rm{ch-eq}} \delta_{j_1 j_2} \delta (x - x') $.
Note that the correlation function is invariant under $SU(3)$ rotations owing to the random mixing of the neutral fields; a derivation can be found in Appendix~\ref{appen_disorderaveraging}.
The renormalization group scaling of the disorder variance $W_{\rm{ch-eq}}$ gives rise to the elastic scattering length $\ell_{\rm{ch-eq}, 0} \sim 1/W_{\rm{ch-eq}}^{3/5}$ (see details in Appendix~\ref{app_length}).  After performing the disorder averaging, an effective  action on the Keldysh contour $K$ reads
\begin{align} \label{effectiveaction}
S_{\textrm{ch-eq}, \textrm{eff}} &= i W_{\rm{ch-eq}} \sum_{j} \int\!  \! dx \! \int_{K} \! dt dt'  e^{i \phi_{1/3} (x, t) }  e^{-i \phi_{c} (x, t) / \sqrt{3} }  \nonumber \\ 
& \times \psi_{j}^{\dagger} (x, t)  \psi_{j} (x, t') e^{i \phi_{c} (x, t') / \sqrt{3}}  e^{-i \phi_{1/3} (x, t') }.
\end{align} 
This action is invariant under the $SU(3)$ transformation $\psi_{j} (x) \rightarrow U(x) \psi_{j} (x)$, unlike the Hamiltonian Eq.~(\ref{qptunnelingeq});
the $SU(3)$ symmetry of the Hamiltonian Eq.~(\ref{bareHamilonian}) is thus restored in a statistical sense. 

We next discuss a discrete transformation $T_3$ defined via
$T_3^{\dagger} \left (\phi_c / \sqrt{3} \right ) T_3 =  \left (\phi_c/\sqrt{3} - 2\pi /3 \right )$,
$T_3^{\dagger} \left (\phi_{n_2}/{\sqrt{2}} \right ) T_3=  \left (\phi_{n_2}/\sqrt{2}  - 2\pi / 3 \right )$, and
$T_3^{\dagger} \phi_{n_1}  T_3 = \phi_{n_1}$.
In the original basis,  $T_3^{\dagger} \phi_4 T_3 = \phi_4 - 2\pi /3$ creates a kink, associated with the annihilation of a charge $e/3$ quasi-particle in $\phi_4$  (cf. Fig.~1(a)). 
Such properties of $T_3$ suggest the explicit form of $T_3 = e^{i \int_{-\infty}^{\infty} dx \partial_x \phi_{4} (x)} $.
In the basis of neutralons 
\begin{align} \label{Discretesymmetry}
T_3^{\dagger} \psi_{j } T_3 = \psi_j e^{2 \pi i /3} \,\,\,\,\,\,\,\,\,  T_3^{\dagger} \psi_{\bar{j} } T_3 = \psi_{\bar{j}} e^{-2 \pi i /3}
\end{align}
for $j = u, d, s$, reflecting the $Z_3$ nature of $T_3$. We find that
$T_3$ is a symmetry of $H_0+ H_{\rm{ch-eq}}$
(Eqs.~\eqref{bareHamilonian} and \eqref{qptunnelingeq}); here a quasi-particles is not allowed to  tunnel into $\phi_4$.

Electron tunneling among the original chiral modes is strong near the intermediate fixed point. Provided that the distance between those chiral modes is not too large, the $\nu=0$ strip (Fig.~\ref{edgest1}(a)) will be smeared, and quasi-particle tunneling to/from $\phi_4$ will be facilitated, reaching the reconstructed edge structure depicted in Fig.~\ref{edgest1}(b). This, in terms of the renormalized modes, will give rise to the following Hamitonian in the neutral sector (assumed to be a small perturbation):
\begin{align} \label{mixing}
H_{n\bar{n}} &=
\int dx (v_{u} \psi_{\bar{d}}^{\dagger} \psi_{s} + v_{d} \psi_{\bar{s}}^{\dagger} \psi_{u}
+ v_{s} \psi_{\bar{u}}^{\dagger} \psi_{d} + \textrm{H.c.}) \ .
\end{align}
This Hamiltonian describes neutralon-antineutralon mixing. 
For the microscopic origin of Eq.~\eqref{mixing}, see Appendix~\ref{appen_tunnelingop}.
In view of Eq.~(\ref{Discretesymmetry}), $H_{n\bar{n}}$  breaks the $Z_3$ symmetry. Here, the annihilation operator of anti-neutralons is defined as $\psi_{\bar{j} = \bar{u}, \bar{d}, \bar{s}} = \psi_{j} ^{\dagger}$.  We note that Eq.~(\ref{mixing}) is the neutralon analogue of a BCS Hamiltonian.  The Hamiltonian, Eq.~(\ref{mixing}), also breaks the $SU(3)$ symmetry. Note that the tunneling amplitudes $v_{j = u, d, s}$ are random with a white noise correlator $\langle v_{j_1} (x) v_{j_2}^{*} (x') \rangle_{\textrm{dis}} = W_{n \bar{n}} \delta (x - x') \delta_{j_1 j_2}$. Note that the tunneling amplitudes $v_{j = u, d, s}$ have the same disorder correlation $W_{n \bar{n}}$ due to the random rotation of the neutral fields by disorder. This allows us to restore the $SU(3)$ symmetry by performing disorder averaging. The characteristic elastic length scale $\ell_{n\bar{n}, 0}$  for this process scales  as 
$\ell_{n\bar{n}, 0} \sim 1/ W_{n \bar{n}}^{3/7}$.

Unless the $Z_3$ is broken, the fundamental representation furnished by neutralons is completely disconnected from its conjugate representation furnished by 
anti-neutralons within fixed charge sectors.  To see this, we consider a correlator  $M_{j \rightarrow \bar{j'}} \equiv \langle 0 | \psi_{\bar{j'}} (x_f, t_f)  \psi_{j}^{\dagger} (x_i, t_i) | 0 \rangle$ between neutralon $j = u, d, s$ at position $x_i$ and time $t_i$ and anti-neutralon $\bar{j'} = \bar{u}, \bar{d}, \bar{s}$ at position $x_f$ and time $t_f$. Here $\psi_{\bar{j'}} (x_f, t_f)$ and  $\psi_{j}^{\dagger} (x_i, t_i) $ are operators in the Heisenberg picture. We assume that $Z_3$ is a symmetry of the system: $[H_0 + H_{\rm{ch-eq}}, T_3] = 0$. Then, the vacuum state $|0 \rangle$ must be an eigenstate of $T_3$, $T_3|0 \rangle = t_0|0 \rangle$.
Employing the symmetry condition $[T_3, H_0 + H_{\rm{ch-eq}}] = 0 $ and Eq.~\eqref{Discretesymmetry}, $M_{j \rightarrow \bar{j'}}$ is shown to be zero since
\begin{align} \label{correlator}
M_{j \rightarrow \bar{j'}} &=  \langle 0 | U^{\dagger} (t_f) \psi_{\bar{j'}} (x_f)  U (t_f - t_i) \psi_{j}^{\dagger} (x_i) U (t_i) | 0 \rangle \nonumber \\ 
&=  \langle 0 | T_3  T_3^{\dagger } U^{\dagger} (t_f) T_3  T_3^{\dagger } \psi_{\bar{j'}} (x_f) T_3  T_3^{\dagger }  U (t_f - t_i) T_3  T_3^{\dagger } 
\nonumber \\  & \times
\psi_{j}^{\dagger} (x_i) T_3  T_3^{\dagger } U (t_i) | 0 \rangle \nonumber \\ 
& = e^{- 4 \pi i /3} M_{j \rightarrow \bar{j'}}.
\end{align}
Here, the operator $T_3^\dagger$  was moved from left to right using the fact that it commutes with the time evolution operator 
 $U(t) = e^{- i (H_0 + H_{\rm{ch-eq}}) t}$, and $|t_0|^2 = 1$ is used. Since Eq.~(\ref{correlator}) can only be satisfied for $M_{j \rightarrow \bar{j'}} =0$, we conclude that  in order to allow mixing between  neutralons and anti-neutralons,  
the  $Z_3$ symmetry should be broken. In the next section, we will explicitly show that the conductance plateau of $e^2/ (3h)$ is achieved only if this is indeed the case.

We next consider the two terminal setup depicted in Fig.~\ref{Process2}; the QPC is set such that the outer-most mode is fully transmitted while the inner modes are fully reflected. Such a QPC configuration naively (neglecting possible contributions from neutalons) gives rise to to the conductance of $e^2/ (3h)$ to drain $\rm{D}$. After transmission through or reflection from the QPC, 
the biased charge modes start to equilibrate with the unbiased ones via quasi-particle tunneling (described by Eq.~\eqref{qptunnelingeq}) in the upper right or lower left corner of Fig.~\ref{Process2}(a). Such tunneling events generate neutralons in the upper right corner or anti-neutralons in the lower left corner, which then 
 propagate in a direction opposite  to that of the charge modes through the QPC region with  size  $L_{\rm{QPC}}$, and
finally  decay in the lower right or upper left corner ("decay region") as depicted in Fig.~\ref{Process2}(b). 
Depending on whether the system exhibits the $Z_3$ symmetry or not, we distinguish between two models: 
model (B) includes the terms of Eq.~\eqref{mixing} to break the $Z_3$, while  model (A) does not.
For a clear contrast between the models, $\ell_{n\bar{n}} \ll L_{\rm{QPC}}$ is assumed for the model (B) to achieve the complete mixing between neutralons and anti-neutralons. 
In both models (A) and (B), we assume that the $SU(3)$ symmetry is statistically conserved, with $\ell_{\rm ch-eq, 0} \ll 
L_{\rm arm}$. 

\section{Transport properties: Tunneling current and Non-equilibrium noise} 

We now turn our attention to transport properties of the two terminal setup: d.c. conductance and non-equilibrium noise. We first demonstrate that the $Z_3$ symmetry of model (A) prevents the formation of  a $e^2/ (3h)$ conductance plateau. 
To understand this, it suffices to consider a single impurity in the decay region in the lower/upper edge $r = \rm{l}/\rm{u}$. The impurity is assumed to be at position $x_0$. Let us start with  model (A). 
To leading order in $t_{j}$, the tunneling current at $x_0$ on the $r = \rm{l}, \rm{u}$ edge can be expressed in terms of a local greater (lesser) Green's function 
$g^{>}_{j, r} (t) = -i \big \langle \psi_{j, r} \left ( x_0, t \right )  \psi_{j, r}^{\dagger} \left ( x_0, 0 \right )  \big \rangle$
($g^{<}_{j, r} (t) = -i \big \langle \psi_{j, r}^{\dagger} \left ( x_0, 0 \right ) \psi_{j, r} \left ( x_0, t \right )    \big  \rangle$) of $j = u, d, s$ type  neutralons as 
\begin{align} \label{tunnelingcurrent}
I_{\rm{tun}, r} = &\sum_{j = u, d, s} \frac{i e |t_j|^2}{3 (ah)^2} \int_{-\infty}^{\infty} dt \left( \frac{a}{a + i v_c t} \right)^{1/3} \nonumber \\ 
& \times \left( \frac{a}{a + i v_{1/3} t} \right)^{1/3} \left [g_{j, r}^{>} (t) - g_{j, r}^{<} (-t) \right ].
\end{align}
Here, $a$ is a short distance cutoff, and $v_{1/3}$ and $v_c$ are the velocity of the outer-most mode and the inner charge mode, respectively. 
Information about the non-equilibrium state of neutralons is encoded in the local greater and lesser Green's function~\cite{Gutman2008, Levkivskyi2009, Gutman2010, Levkivskyi2012, Rosenow2016, Levkivskyi2016} via the decomposition 
\begin{align} \label{correlationFun}
g^{>}_{j, r} \left ( t\right ) = g_0 (t) g_{r}^{\rm{neq}} (t), \,\,\,\,\,\,\,\,
g^{<}_{j, r} \left ( t\right ) =g_0 (-t) g_{r}^{\rm{neq}} (t) \ .
\end{align}
 Here,  $g_{0} (t) =  -i \left [ a/ (a + i v_n t) \right ] ^{2/3}$ describes quantum correlations of neutralons, while $g_{r}^{\rm{neq}} (t)$ represents classical  non-equilibrium aspects of neutralons, as
\begin{align} \label{noneqGreen}
g_{r, \rm model \, A}^{\rm{neq}} (t)  = \left[ \frac{1}{2} + \frac{1}{2} e^{2 \pi i r \textrm{sgn} (t)/3 }\right ]^{N}
\end{align}
with  $N = e V |t| / \hbar$. We briefly sketch the derivation of Eq.~(\ref{correlationFun}). In the limit of full charge equilibration,  quasi-particles emanating from source S and then being transmitted through the QPC towards drain D, reach the charge mode with probability $1/2$. This is accompanied by the creation of neutralons with probability $1/2$, 
cf.~Eq.~(\ref{noneqGreen}). 
Each of these neutralons arrives  at $x_0$, described by a kink in the bosonic fields 
$\phi_{n_1}(x_0)$ and $\phi_{n_2}(x_0 )$, 
and thus gives rise to a phase shift  $2 \pi\, \textrm{sgn} (t) /3 $ of the operator
$\psi_j (x_0, t) \psi_j^{\dagger}(x_0, 0)$: following the arrival of a $j' = u, d, s$ neutralon at $x_0$ and time $t'$,
$\psi_j (x_0, t) \psi_j^{\dagger}(x_0, 0)$ acquires a phase shift of  $2 \pi  \textrm{sgn} (t) /3 $.  This relies on
\begin{align} \label{phaseshift}
& \psi_{j'} (x_0, t') \left [\psi_j (x_0, t) \psi_j^{\dagger}(x_0, 0) \right ] \psi_{j'}^{\dagger} (x_0, t') \nonumber \\ = & \left [\psi_j (x_0, t) \psi_j^{\dagger}(x_0, 0) \right ] e^{ \pi i\left [ \textrm{sgn} (t- t') - \textrm{sgn} (0- t') \right ] /3 },
\end{align}
provided $\textrm{min} (0, t) < t' < \textrm{max} (0, t) $. 
For each arriving neutralon, the phase factor on the r.h.s. of Eq.~(\ref{phaseshift})  appears in one of the $N$ factors   of Eq.~\eqref{noneqGreen}.
We note that the phase factor reflects the anyonic statistics of neutralons and is identical for all flavors $j' = u, d, s$.
When no  neutral excitation is generated (with a probability $1/2$) during the equilibration process, $\psi_j (x_0, t) \psi_j^{\dagger}(x_0, 0)$ does not accumulate a phase factor, leading to the first term
in the parenthesis of Eq.~\eqref{noneqGreen}. 
Inserting Eqs.~\eqref{correlationFun} and \eqref{noneqGreen} into Eq.~\eqref{tunnelingcurrent}, we obtain  the tunneling current as
\begin{align} \label{tunnelingcurrentfinal}
I_{\rm{tun}, r} = \sum_{j = u, d, s} \frac{2 r c e  |t_j|^2}{3 \hbar \Gamma (\frac{4}{3})}
\left (\frac{eV}{a^2 h^4 v_n^2 v_c v_{1/3}} \right )^{\frac{1}{3}},
\end{align}
where $c = \sin \left [ \tan^{-1} \left [\pi / (3 \ln (2)) \right ] /3 \right ] \left [
(\pi/3)^2 + (\ln (2))^2\right ]$ and $\Gamma (x) $ is the gamma function. 
The finite value of the tunneling current causes a deviation of the conductance from $e^2 / (3h)$.
If several impurities are taken into account and all the neutral excitations eventually decay ($\ell_{\rm{ch-eq, 0}} \ll L_{\rm{arm}}$), 
it can be self-consistently shown that the conductance between source and drain is zero up to an exponential correction $\sim \exp(-L_{\rm{arm}}/ \ell_{\rm{ch-eq, 0}})$ (see Appendix \ref{appen_z3sym} for a derivation).  

In the framework of  model (B), the full mixing between neutralons and anti-neutralons in the QPC region ($\ell_{n\bar{n}, 0} \ll L_{\rm{QPC}}$) causes
both types of particles to arrive with the same probability at $x_0$. As the phase shifts of neutralons and anti-neutralons are the complex conjugate of each other, the non-equilibrium part of the Green's functions 
\begin{align} \label{noneqGreenmodelB}
g_{r, \rm model \,  B}^{\rm{neq}} (t)  = \left[ \frac{1}{2} + \frac{1}{4} e^{2 \pi i  \textrm{sgn} (t)/3 } + 
\frac{1}{4} e^{-2 \pi i  \textrm{sgn} (t)/3 }
\right ]^{N}.
\end{align}
is real and leads to a vanishing tunneling current when inserted into 
 Eqs.~\eqref{tunnelingcurrent}-\eqref{correlationFun}, causing the conductance quantization of $e^2/(3h)$. The incomplete mixing of neutralons with anti-neutralons leads to an exponential correction $\propto \exp (- L_{\rm{QPC}}/ \ell_{n \bar{n},0})$ of the quantized value.

\begin{table*}
\centering
\begin{tabular}[c]{ | c || c | c | c|  } 
\hline
& Exact $SU(3)$ symmetry & $\ell_{\rm{ch-eq, 0}} \gg L_{\rm{arm}}$ & $\ell_{\rm{ch-eq, 0}} \ll L_{\rm{arm}}$ \\
 \hline
\multirow{2}{*}{Conserved $Z_3$ symmetry}
 & Conductance plateau $e^2 / (3h)$   & Mesoscopic conductance fluctuation &   Zero conductance\\ 
  & Zero noise on the plateau  & $G= e^2 (1 - R) / (3h)$, $S_{\rm{D}}=  2e^3 R V_0 / (9h)$& Zero noise\\
\hline
Broken $Z_3$ symmetry & \multirow{2}{*}{\diagbox[width=43mm]{~}{~}} &  Conductance plateau $e^2 / (3h)$   &  Conductance plateau $e^2 / (3h)$\\
($\ell_{n\bar{n}, 0} \ll L_{\rm{QPC}} $)& &  Non-universal noise & Fano factor of $2e/3$ \\
 \hline
\end{tabular}
\caption{A summary of experimental manifestations of two symmetries, a continuous one $SU(3)$ and a discrete threefold one $Z_3$ (cf. Eq.~\eqref{Discretesymmetry}).
 Model (A) (model (B)) corresponds to the cell in the first (second) row and the third column. 
}
\label{table1}
\end{table*}

We now quantify the zero-frequency non-equilibrium noise measured at the drain for model (B). 
Using the non-equilibrium bosonization technique~\cite{Gutman2008, Levkivskyi2009, Gutman2010, Levkivskyi2012, Rosenow2016, Levkivskyi2016}, we compute generating functions
 $g_{n_1} = \big \langle  e^{i  \lambda_1 \phi_{n_1} (x, t)} e^{-i  \lambda_1 \phi_{n_1} (x', t')} \big  \rangle$ and $g_{n_2} =  \big \langle  e^{i  \lambda_2 \phi_{n_2} (x, t)} e^{- i  \lambda_2 \phi_{n_2} (x', t')} \big \rangle$ as 
\begin{align} \label{generatingFun1}
g_{n_1} &= 
 \left( \frac{a}{a + i v_n \tau  }\right)^{\lambda_1^2 }\left[ \frac{2}{3}   
+  \frac{\cos \left (\pi\sqrt{2} \lambda_1 \right)}{3} \right ]^{M},  \\  \label{generatingFun2}
g_{n_2} &= 
 \left( \frac{a}{a + i v_n \tau }\right)^{\frac{\lambda_2^2 }{3}}
 \left[ \frac{1}{2}   
+  \frac{\cos \left( \frac{\sqrt{2} \pi   \lambda_2}{ 3} \right ) }{3} + \frac{\cos \left( \frac{2\sqrt{2} \pi   \lambda_2}{ 3} \right ) }{6} \right ]^{M},
 \end{align} 
valid in the long time limit $M \equiv eV |\tau| / h \gg 1$, with   $\tau \equiv  (t -t')+ (x - x')/ v_n$. 
To derive Eqs.~(\ref{generatingFun1}) and (\ref{generatingFun2}), we have used that up and down neutralons  create a kink of height $\pm \sqrt{2} \pi$ in $\phi_{n1}$, while the strange neutralons leave $\phi_{n1}$ invariant. Likewise, 
 up and down neutralons create a kink of height of $+\sqrt{2} \pi /3 $ in $\phi_{n_2}$, while strange neutralons
create a kink of height  $-2 \sqrt{2} \pi /3 $.

By taking second derivatives, we obtain the correlation functions $C_{n_i} = \big \langle \phi_{n_i} (x, t) \phi_{n_i} (x', t') \rangle -\langle \phi_{n_i}^2 (x, t) \big \rangle$ ($i=1,2$ labels the two neutral modes) as
\begin{align} \label{correlators}
C_{n_1} (\tau) &= \ln \left( \frac{a}{a + i v_n \tau  }\right)- \frac{eV | \tau| \pi^2  }{3h}, \nonumber \\ 
C_{n_2} (\tau) &=\frac{1}{3} \ln \left( \frac{a}{a + i v_n \tau  }\right)- \frac{eV | \tau| \pi^2  }{9h}. 
\end{align}
The noise in the neutral currents $I_{n_1} (x, t) = v_n \partial_{x} \phi_{n_1} / (\sqrt{2} \pi )$ and 
$I_{n_2} (x, t) = 3v_n \partial_{x} \phi_{n_2} / (\sqrt{2} \pi )$  evaluated in the decay region of edge $r = \textrm{u}/\textrm{l}$
is defined as $S_{n_{i = 1, 2}, r} \equiv 2 \int dt \langle I_{n_1} (x_{r}, t )I_{n_1} (x_{r}, 0) \rangle$. 
Employing Eq.~\eqref{correlators}, we obtain 
\begin{align} \label{noiseNeutralmode}
S_{n_{1},\textrm{u}/ \textrm{l}} &= \frac{v_n^2}{\pi^2} \int_{-\infty}^{\infty} dt \partial_x \partial_{x'} C_{n_1} (\tau)|_{x=x', t'=0} \nonumber \\
&=   - \frac{1}{\pi^2} \int_{-\infty}^{\infty} \partial_\tau^2 C_{n_1} (\tau) d\tau = \frac{2eV}{3h},
\nonumber \\
S_{n_{2},\textrm{u}/ \textrm{l}}&= \frac{2eV}{h}.
\end{align}
We emphasize that even though the second term of Eq.~\eqref{correlators} only holds for the large $|\tau|$ limit, Eq.~\eqref{noiseNeutralmode} 
is valid because it only depends on the values of $\partial_{\tau} C_{n_1}$ and $\partial_{\tau} C_{n_2}$ at $\tau \rightarrow \pm \infty$;
it is insensitive to details of  $\partial_{\tau} C_{n_1}$ and  $\partial_{\tau} C_{n_2}$ at small $\tau$.
Due to the decay of all neutral excitations, the electrical noise measured at drain D can be expressed as
\begin{align} \label{noiserelation}
S_{\textrm{D}}= \frac{e^{2}}{36} (3 S_{n_1, \textrm{u}} +  S_{n_2, \textrm{u}}
+ 3 S_{n_1, \textrm{l}} + S_{n_2, \textrm{l}}) \ .
\end{align}
We refer the reader to Appendix~\ref{appen_noise} for a derivation of Eq.~\eqref{noiserelation}. 
Employing Eqs.~\eqref{noiseNeutralmode} and \eqref{noiserelation}, we obtain the Fano factor  characterizing the strenght of  noise at D as
%
\begin{equation} \label{Fanofactor.eq}
F \equiv  S_{\rm{D}}/ (2 I_{\textrm{imp}} T(1-T)) = 2e/3 \ .
\end{equation}
%
Here, we used that the current impinging on the QPC is given by $I_{\textrm{imp}} = 2e^2 V / (3h)$, and that the   transmission probability through the QPC is $T = 1/2$. 

Experimental manifestations (conductance and dc noise) of possible symmetry scenarios beyond those of model (A) and (B) are summarized in Table.~\ref{table1}. For example, 
when  $Z_3$ is conserved and the $SU(3)$ symmetry is exact (intermediate $SU(3)$-symmetric fixed point), the 
the conductance is quantized at $e^2 / (3h)$ with vanishing noise (there is no tunneling between the charge modes as $\ell_{\rm{ch-eq}, 0} \rightarrow \infty$). 
In yet another scenario $\ell_{\rm{ch-eq}, 0}$ becomes finite but still $\ell_{\rm ch-eq, 0} \ll L_{\rm arm}$. Then,  some quasi-particles emanating from the source are reflected via the two-step mechanism of   neutralon creation followed by the decay of these neutralons, leading to a reduction of the quantized conductance. While mesoscopic fluctuations of the conductance~\cite{Rosenow2010, Protopopov2017} occur when $\ell_{\rm{ch-eq, 0}} \gg L_{\rm{arm}}$, both  conductance and  noise become zero when $\ell_{\rm{ch-eq, 0}} \ll L_{\rm{arm}}$.
When the $Z_3$ is broken (as $\ell_{n\bar{n}, 0} \ll L_{\rm{QPC}}$), on the other hand, the conductance is always  quantized as $e^2 / (3h)$.  The Fano factor however  deviates from the non-universal value to $2e/3$ as $\ell_{\rm{ch-eq, 0}}$ decreases. 
This non-universal Fano factor can be attributed to a partial decay of  neutral excitations, producing a smaller noise as compared to the case of  full decay. The crossover of the transport properties between
$\ell_{\rm{ch-eq, 0}} \gg L_{\rm{arm}}$ and $\ell_{\rm{ch-eq, 0}} \ll L_{\rm{arm}}$ might be experimentally tuned by varying either  $L_{\rm{arm}}$ directly  or the slope of the edge confining potential
to effectively control $\ell_{\rm{ch-eq, 0}}$.

\section{Summary}
For general hole-conjugate quantum Hall states at filling factor $\nu = -n/(n p +1)$, with 
negative even integer $p$ and positive integer $n$, the relevant symmetries to be considered are a $SU(n+1)$  in the neutralon sector and a $Z_{n+1}$ symmetry connecting neutralons to anti-neutralons. Similarly to the $2/3$ case, we expect 
different symmetry scenarios, including a quantized noise on a conductance plateau for statistically preserved $SU(n+1)$ and broken $Z_{n+1}$.

To summarize, we have investigated the roles of two symmetries, a continous $SU(3)$ and a discrete one $Z_3$, in influencing the 
two-terminal conductance and the dc noise of a quantum Hall strip at  filling factor $\nu = 2/3$ with a single QPC.
While  recent measurements~\cite{Bid2009, Sabo2017} with a Fano factor $2e/3$ on the conductance plateau of $e^2 / (3h)$
are explained relying on a broken  $Z_3$ symmetry, other symmetry scenarios (summarized in Table.~\ref{table1}) can be realized in future experiments. 


\begin{acknowledgments}
We thank Amir Rosenblatt, Rajarshi Bhattacharyya,
Ron Sabo, Itamar Gurman, Christian Sp\r{a}nsl\"{a}tt, and Moty Heiblum for helpful discussions. We are indebted to Alexander Mirlin for very useful comments on the manuscript.
B.R. acknowledges support  from the Rosi and Max Varon Visiting Professorship at the Weizmann Institute of Science, B.R. and Y.G. acknowledge support by DFG Grant
No. RO 2247/11-1. Y.G. further acknowledges support from CRC 183 (Project C01), the Minerva Foundation, and   DFG Grant
No. MI 658/10-1.
J.P. acknowledges support by the Koshland Foundation and funding by the Deutsche Forschungsgemeinschaft
(DFG, German Research Foundation)—Projektnummer
277101999—TRR 183 (project A01).
\end{acknowledgments}

\appendix


\section{SU(3) group structure} \label{appen_SU3}

In this section, we consider the intermediate fixed point vis-a-vis the $SU(3)$ group structure. 
As seen in Eq.~\eqref{Neutralsector}, the intermediate $SU(3)$ symmetric fixed point is given by
\begin{widetext}
 \begin{align} \label{appen_Neutralsector}
 S_{n} &=\int dx dt \bigg [ \frac{1}{4\pi} \left (\partial_x \phi_{n_1} (\partial_{t}-v_{n}\partial_x ) \phi_{n_1}
+3\partial_x \phi_{n_2} (\partial_{t}-v_{n}\partial_x ) \phi_{n_2} \right ) \nonumber \\ &
-\frac{1}{2\pi a}\left (\xi_{1}(x) e^{i \sqrt{2} \phi_{n_1}}+ \xi_{2}(x) e^{- i \phi_{n_1} /\sqrt{2}}  e^{3 i \phi_{n_2}/\sqrt{2}} 
+\xi_{3}(x)  e^{ i \phi_{n_1}/\sqrt{2}}  e^{3 i \phi_{n_2} /\sqrt{2}}  + \textrm{H.c.} \right ) \bigg ].
 \end{align}
 \end{widetext}
The second line originates from electron tunneling between the inner modes; $\xi_{i = 1, 2, 3}$ are random variables. It is worthwhile to note the group structure of Eq.~\eqref{appen_Neutralsector}.
 \begin{figure} 
\includegraphics[width=0.7\columnwidth]{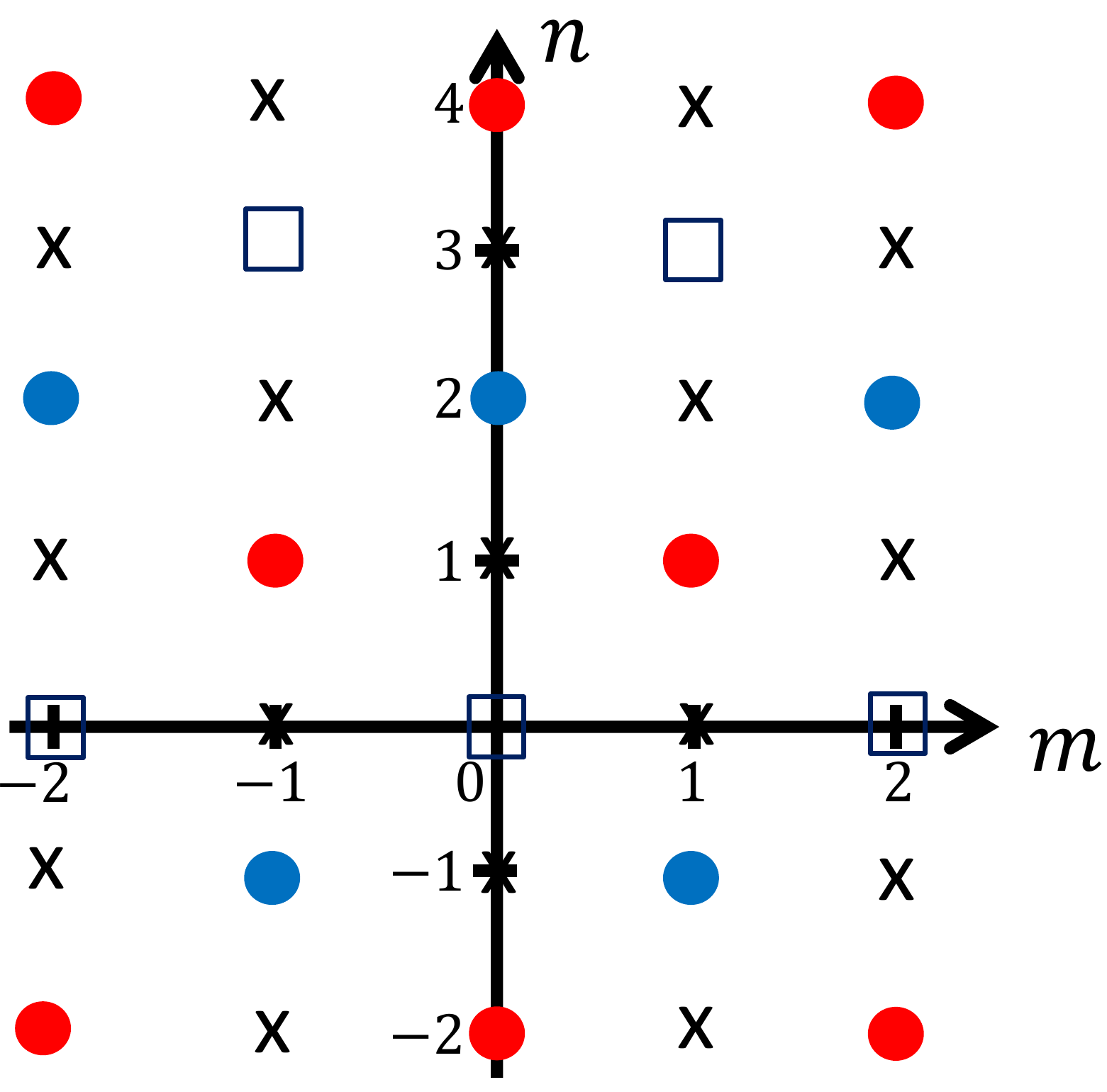} 
\caption{ (color online). A lattice for $SU(3)$ group representations with allowed values of $m$ and $n$ for states of $e^{i m \phi_{n_1} / \sqrt{2}}e^{i n \phi_{n_2} / \sqrt{2}} |0\rangle$. Here $|0\rangle$ is the vacuum state. The points denoted as $\times$ are not allowed. The states are divided into three sub-lattices: the neutralon sector (denoted as red dots), the anti-neutralon sector (denoted as blue dots), and the vacuum sector (denoted as rectangulars). Points within each sub-lattice are connected via the electron operators between the inner modes (written in the second line of Eq.~\eqref{Neutralsector}), while a sub-lattice is completely decoupled from the other sublattices due to Eq.~\eqref{Neutralsector}. The presence of three different sectors manifests the $Z_3$ subgroup of the $SU(3)$ group. 
}\label{SU3Sym}
\end{figure}
A lattice for $SU(3)$ group representations with allowed values of $m$ and $n$ for states of $e^{i m \phi_{n_1} / \sqrt{2}}e^{i n \phi_{n_2} / \sqrt{2}} |0\rangle$ is displayed in Fig.~\ref{SU3Sym}. Here $|0\rangle$ is the vacuum state.  The points denoted as $\times$ are not allowed. The states are divided into three sub-lattices: the neutralon sector (denoted as red dots), the anti-neutralon sector (denoted as blue dots), and the vacuum sector (denoted as rectangulars). Points within each sub-lattice are connected via the electron operators between the inner modes (written in the second line of Eq.~\eqref{appen_Neutralsector}), while a sub-lattice is completely decoupled from the other sublattices by the electron tunneling. The presence of the three disconnected sectors manifests the $Z_3$ subgroup of the $SU(3)$ group. Importantly, it should be noted that the neutralon sector is completely disconnected from the anti-neutralon sectors by the $SU(3)$ symmetry. For the connection between the neutralon and the anti-neutralon sector, the discrete symmetry $T_3$ (defined in Eq.~\eqref{Discretesymmetry}) should be broken.  
 
Including the action for an auxiliary field $\chi$ (see Ref.~\cite{Kane1994} for a similar procedure)
and performing refermionization in terms of a three-component fermion field $\Psi (x) =  e^{i \chi /\sqrt{3}}  [e^{-i(\phi_{n_1}+\phi_{n_2})/\sqrt{2}}, e^{i(\phi_{n_1}-\phi_{n_2})/\sqrt{2}}, e^{i\sqrt{2}\phi_{n_2}}]^T /\sqrt{2 \pi a} = e^{i \chi / \sqrt{3}} [\psi_{u}, \psi_{d}, \psi_{s}]^T /\sqrt{2 \pi a}$, Eq.~\eqref{appen_Neutralsector} reads
\begin{widetext}
  \begin{align} \label{Appen_Neutralsector1}
 S_{n} &= \int dx dt \left [ i\Psi^{\dagger} (-\partial_t +v_n \partial_x) \Psi - \left ( \xi_{1} \Psi^{\dagger} (\frac{\lambda_1 + i \lambda_2}{2}) \Psi + \xi_{2} \Psi^{\dagger} (\frac{\lambda_6 + i \lambda_7}{2}) \Psi+ \xi_3 \Psi^{\dagger} (\frac{\lambda_4 + i \lambda_5}{2})\Psi + \textrm{H.c.} \right ) \right].
 \end{align}
 $\lambda_{1,2,4,5,6,7}$ are non-diagonal generators of the $SU(3)$ group, given by 
\begin{align} \label{SU3generator}
\lambda_1 = \begin{pmatrix}
      0 & 1 & 0\\
   1 & 0 & 0 \\
  0  & 0 & 0  
 \end{pmatrix},\,\,\,\,\,\,\, 
\lambda_2 = \begin{pmatrix}
      0 & -i & 0\\
   i & 0 & 0 \\
  0  & 0 & 0  
 \end{pmatrix},\,\,\,\,\,\,\, 
\lambda_4 = \begin{pmatrix}
      0 & 0 & 1\\
   0 & 0 & 0 \\
   1 & 0 & 0  
 \end{pmatrix}, \nonumber \\ 
\lambda_5 = \begin{pmatrix}
      0 & 0 & -i\\
   0 & 0 & 0 \\
   i & 0 & 0  
 \end{pmatrix},\,\,\,\,\,\,\,
\lambda_6 = \begin{pmatrix}
      0 & 0 & 0\\
  0 & 0 & 1 \\
  0  & 1 & 0  
 \end{pmatrix},\,\,\,\,\,\,\,
\lambda_7 = \begin{pmatrix}
      0 & 0 & 0\\
   0 & 0 & -i \\
   0 & i & 0  \end{pmatrix}. 
\end{align} 
\end{widetext}
The diagonal generators $\lambda_3$ and $\lambda_8$ are associated with the density of the neutral modes, $
\partial_x \phi_{n_1}$ and  $
\partial_x \phi_{n_2}$, and are explicitly written in Eq.~\eqref{diagonalgenerator}.
 The random terms are completely eliminated via performing a $SU(3)$ gauge transformation of 
 $\tilde{\Psi} (x) = U(x) \Psi (x) $; the action becomes diagonal as $S_{n} = i \int dx dt \tilde{\Psi}^{\dagger} (-\partial_t + v_n \partial_x ) \tilde{\Psi}$. The position dependent $SU(3)$ matrix $U(x)$ is given in the main text (Eq.~\eqref{rotationmatrix}).

\section{Tunneling operators} \label{appen_tunnelingop}
In this section, we derive the form of quasi-particle tunneling operators in terms of neutralons.

We first consider quasi-particle tunneling between the inner modes and the outermost mode at the vicinity of the intermediate fixed point, given by
\begin{widetext}
\begin{align} \label{appen_qptunnelingeq}
H_{\rm{ch-eq}}& =\int dx \left [t_{u, 0}  e^{ i (\phi_{1/3} + \phi_2)}  +  t_{d, 0}  e^{i(\phi_{1/3}- 2\phi_2- \phi_3)} 
+ t_{s, 0} e^{i(\phi_{1/3}- 2 \phi_2 - 2\phi_3 -3\phi_4)}  + \textrm{H.c.} \right ]
\nonumber \\ & =   \int dx \left [ (t_{u} \psi_{u}^{\dagger}+ t_{d}  \psi_{d}^{\dagger}  
+ t_{s}  \psi_{s}^{\dagger}) e^{- i \phi_c /\sqrt{3}} e^{ i\phi_{1/3}} + \textrm{H.c.} \right ].
\end{align}
\end{widetext}
All the tunneling operators have the same scaling dimension of $\delta_{\textrm{ch-eq}} = 2/3$. Bare tunneling amplitudes $t_{j, 0}$ are random with white noise correlation $\langle t_{j_1, 0} (x) t_{j_2, 0}^{*} (x') \rangle_{\textrm{dis}} = W_{\textrm{eq}, j_1}^0 \delta (x - x') \delta_{j_1 j_2}$. In the rotated basis of the $\psi_{j}$, the tunneling amplitudes $t_{j=u,d,s}$ are given by
$t_{j} = \sum_{j'} U_{j j'} t_{j', 0}$.

The other type of quasi-particle tunneling (related to quasi-particle tunneling to the $\phi_4$ mode) occurs when the electron tunneling between the inners is strong enough to eliminate the regions of the filling factors $\nu= 1$ and $\nu = 0$, eventually reaching the edge structure of Fig.~\ref{edgest1}(b). The most relevant terms are written as 
\begin{widetext}
\begin{align}\label{appen_qptunnelingmixing}
H_{n \bar{n}}& =  \int dx \left (v_{u, 0} e^{i (2\phi_2 + \phi_3 + \phi_4)} + v_{d, 0}e^{- i (\phi_{2} - \phi_{4})}  
+ v_{s, 0} e^{-i (\phi_2 + \phi_3 + 2\phi_4)} + \textrm{H.c.} \right )  \nonumber \\ &=
\int dx \left (v_{u} \psi_{\bar{d}}^{\dagger} \psi_{s} + v_{d} \psi_{\bar{s}}^{\dagger} \psi_{u}
+ v_{s} \psi_{\bar{u}}^{\dagger} \psi_{d} + \textrm{H.c.} \right). 
\end{align}
\end{widetext}
All the terms have the same scaling dimension of $\delta_{n \bar{n}} = 1/3$. 
Bare tunneling amplitudes $v_{j, 0}$ are random with white noise correlation $\langle v_{j_1, 0} (x) v_{j_2, 0}^{*} (x') \rangle_{\textrm{dis}} = W_{n\bar{n}, j_1}^0 \delta (x - x') \delta_{j_1 j_2}$.

\section{SU(3) symmetry restoration by disorder averaging} \label{appen_disorderaveraging}
In this section, it is shown that after the self-averaging over the disorder, the effective action for the quasi-particle tunneling becomes invariant under the $SU(3)$ transformation. 

To this end, we consider the quasi-particle tunneling (Eq.~\eqref{appen_qptunnelingeq}) between the outer-most mode and the inner ones. The tunneling amplitudes $t_{j=u,d,s}$ is random with the noise correlation as
\begin{align} \label{disorder}
\langle t_{j_1} (x) t_{j_2}^{*} (x') \rangle_{\textrm{dis}} =& \sum_{j, j'} \left \langle U_{j_1j} (x) t_{j, 0} (x) U_{j_2, j'}^{*} (x') t_{j', 0} (x') \right \rangle_{\textrm{dis}}  \nonumber \\ 
=& \sum_{j, j'} \left \langle U_{j_1j} (x) U_{j_2, j'}^{*} (x')  \right \rangle_{\textrm{dis}} \nonumber \\ & 
\times \left \langle t_{j, 0} (x) t_{j', 0} (x') \right \rangle_{\textrm{dis}}  \nonumber \\ 
=& \sum_{j} \left \langle U_{j_1j} (x) U_{j_2, j}^{*} (x')  \right \rangle_{\textrm{dis}}   W_{\textrm{eq}, j}^0 \delta (x - x')  \nonumber \\ =  & U_0 \delta (x-x') \delta_{j_1, j_2} \sum_{j} W_{\textrm{eq}, j}^{0} \nonumber \\ = &
W_{\rm{ch-eq}} \delta (x-x') \delta_{j_1, j_2}. 
\end{align}
 Here the second equality in Eq.~\eqref{disorder} is due to the statistical independence of $U$ and $t_0$ with regards to the disorder averaging and 
$ \sum_{j} \langle U_{j_1j} (x) U_{j_2, j}^{*} (x')  \rangle_{\textrm{dis}} = U_0 \delta_{j_1, j_2} \delta (x-x')$ comes from the randomness of $U$. 
We defined $W_{\rm{ch-eq}}\equiv U_0  \sum_{j} W_{\rm{eq}, j}^0$.
Eq.~\eqref{disorder} shows that the variance of the tunneling amplitudes in the rotated basis does not depend on the neutralon flavor. This claim is also applicable to the variances of the tunneling amplitudes for $H_{n \bar{n}}$ (Eq.~\eqref{mixing}) through a similar procedure. 

Employing this claim, we show that  after the self-averaging over the disorder, the effective action becomes invariant under the $SU(3)$ transformation. To this end, we consider the corresponding Keldysh action written as 
\begin{align} \label{Keldyshaction}
S_{\rm{ch-eq}} = - \int dx \int_K dt & \big [ (t_{u} \psi_{u}^{\dagger}+ t_{d}  \psi_{d}^{\dagger}  
+ t_{s}  \psi_{s}^{\dagger}) e^{- i \phi_c /\sqrt{3}} e^{ i\phi_{1/3}}
\nonumber \\  &+ \textrm{c.c.} \big ] \ .
\end{align}
The disorder averaging of $e^{iS_{\rm{ch-eq}}}$ can be easily performed by 
\begin{widetext}
\begin{align}
\left \langle e^{iS_{n\bar{n}}}  \right \rangle_{\textrm{dis}} = \frac{\int Dt_j Dt_j^{*}
D\psi_j D\psi_{j}^{\dagger} \exp \left [- \sum_{j}\int dx t_j (x) t_j^{*} (x) / W_{\rm{ch-eq}} + i S_{n\bar{n}}\right ]}{\int Dt_j Dt_j^{*}
 e^{- \sum_{j}\int dx t_j (x) t_j^{*} (x) / W_{\rm{ch-eq}}}}.
\end{align}
\end{widetext}
The integration of $\left \langle e^{iS_{n\bar{n}}}  \right \rangle_{\textrm{dis}} \equiv \int 
D\psi_j D\psi_{j}^{\dagger} e^{i S_{\rm{ch-eq}, \rm{eff}}}$ over the random amplitudes leads to the effective action Eq.~\eqref{effectiveaction}.
This effective action is invariant under the $SU(3)$ transformation $\psi_{j} (x) \rightarrow \sum_{j'} U_{j j'}(x) \psi_{j'} (x)$ unlike the original Hamiltonian $H_{\rm{ch-eq}}$.
Thus the $SU(3)$ symmetry is restored in the statistical sense. 

\section{length scales} \label{app_length}
\begin{figure} 
\includegraphics[width=0.8\columnwidth]{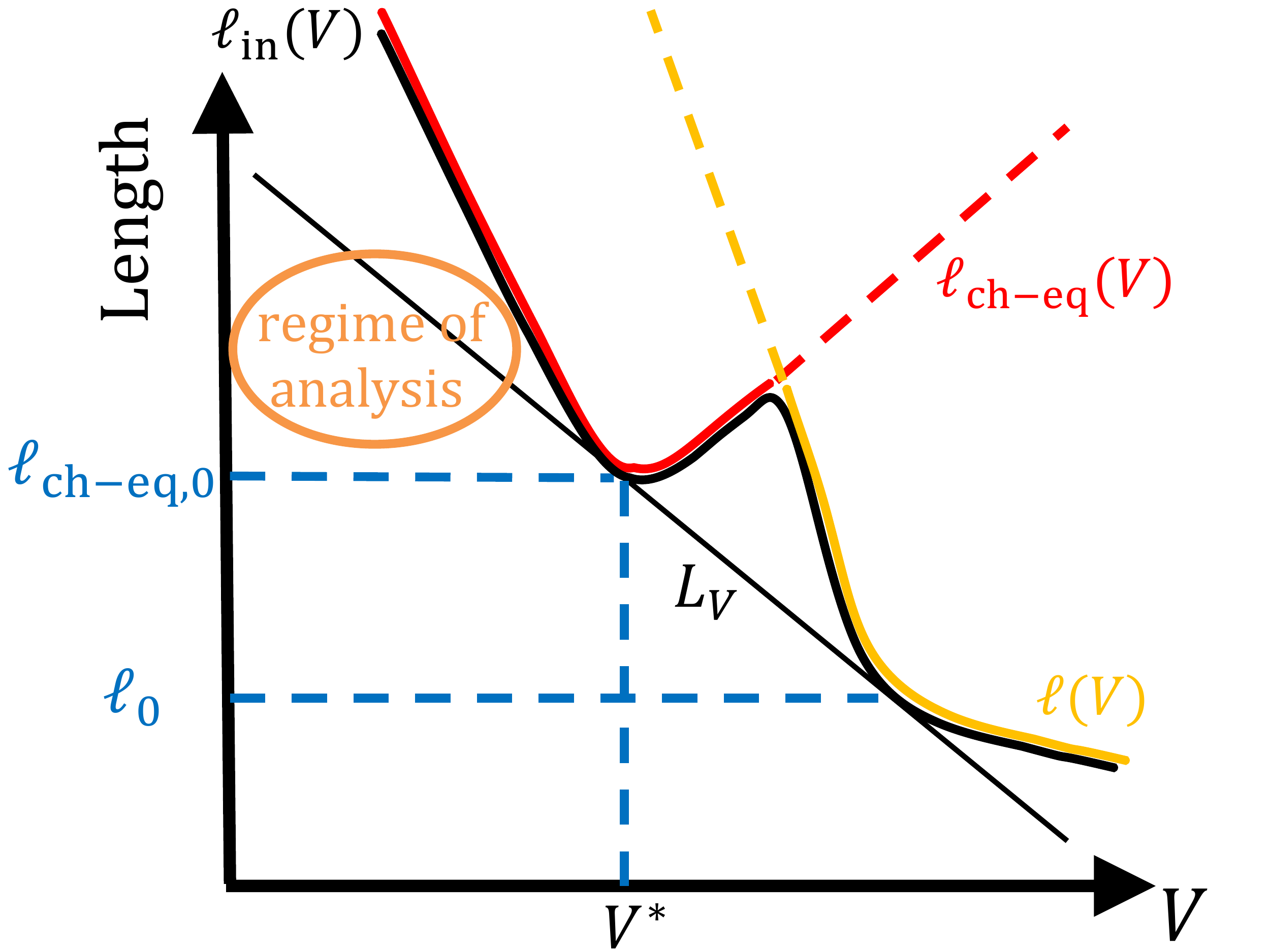} 
\caption{ (color online). The schematic dependence of the relevant length scales on the applied voltage $V$ (log-log plot);
$\ell_{0}$ and $\ell_{\rm{ch-eq, 0}}$ (blue dashed lines) are the voltage-independent elastic lengths.
They correspond to elastic scattering within the inner modes and between the inner and outer modes, respectively. The voltage length $L_{V} \propto 1/V$ (black thin line) acts as an infrared cutoff. 
$\ell (V)$ (yellow curve) and $\ell_{\rm{ch-eq}} (V)$ (red curve) are the voltage-dependent inelastic lengths associated with inelastic scattering within the inner modes and between the inner and outer modes, respectively. Specifically, the scaling of $\ell_{\rm{ch-eq}}$ changes from $\ell_{\rm{ch-eq}} \propto 1/V^2$ (low voltage) to $\ell_{\rm{ch-eq}} \propto V^{2/3}$ (high voltage) at the transition voltage $V = V^{*}$. Here the transition voltage $V^{*}$ is defined as the voltage to satisfy $ L_V (V^{*}) = \ell_{\rm{ch-eq, 0}}$. The inelastic length $\ell_{\textrm{in}} (V)$ (at which the modes lose coherence) takes the smaller value between $\ell$ and $\ell_{\rm{ch-eq}}$. Our main focus in this manuscript is on the regime that $L_{\rm{arm}}$ is much larger than $\ell_{\rm{ch-eq, 0}}$, but smaller than $\ell_{\rm{ch-eq}}$. The voltage $V$ is perfectly interchangeable with the temperature $T$ when $T \gg V$.
}\label{lengthscales}
\end{figure}

Here we describe the relevant length scales, and discuss their scaling with an external voltage. We analyze it employing the RG method of Ref.~\cite{Protopopov2017}. The schematic dependence of the relevant length scales on the voltage is depicted in Fig.~\ref{lengthscales}.
We focus on the dependence on the voltage, but the voltage $V$ is perfectly interchangeable with the temperature $T$ when $T \gg V$.

We first summarize all length scales considered in this work: 
\begin{itemize}
          \item $L_{\rm{arm}}$ : the length of the arms between the contacts and the QPC.
          \item $L_{\rm{QPC}}$ : the size of the QPC. 
	\item $\ell_0$ : the elastic scattering length beyond which disorder-induced electron tunneling mixes the inner modes. When $L_{\rm{arm}}$ exceeds $\ell_0$, 
                      disorder becomes relevant and the system is driven to the intermediate fixed point~\cite{Wang2013}. The intermediate fixed point is called the Wang-Meir-Gefen (WMG) fixed point in this section. 
	\item $\ell_{\rm{ch-eq, 0}}$ : the elastic scattering length beyond which disorder-induced quasi-particle tunneling mixes the inner modes with the outer-most mode. 

         \item $\ell_{n\bar{n}, 0}$ : the elastic scattering length over which neutralons are mixed with anti-neutralons.
         \item $\ell_{\rm{ch-eq}}$ : the inelastic scattering length (the red curve in Fig.~\ref{lengthscales})  over which the charge modes are equilibrated. 
        \item $\ell_{n \bar{n}}$ : the inelastic scattering length over which neutralons are equilibrated with anti-neutralons. 
        \item $\ell$ : the inelastic scattering length (the yellow curve in Fig.~\ref{lengthscales}) over which the inner three modes equilibrate. 
        \item $\ell_{\rm{in}}$ : the coherence length (the black thin curve in Fig.~\ref{lengthscales}) at which the modes lose the coherence. It takes the smaller value between $\ell$ and $\ell_{\rm{ch-eq}}$. This can be, in principle, calculated from the analysis of the Boltzmann kinetics. 
         \item $L_{V}$ : the voltage length $\sim 1/V$ (the black thick curve in Fig.~\ref{lengthscales}) operating as an infrared cutoff. 
\end{itemize}

We assume that the electron tunneling operators (Eq.~\eqref{Neutralsector}) within the inner modes drive the system to the WMG intermediate fixed point~\cite{Wang2013}; the length scale $L_{\rm{arm}}$ and $L_{\rm{QPC}}$ 
are much larger than $\ell_{0}$ over which the inner modes are strongly mixed by disorder.
In the vicinity of the WMG fixed point, we consider quasi-particle tunneling $H_{\rm{ch-eq}}$ between the outer-most mode
and the inner modes (Eq.~\eqref{qptunnelingeq}) and neutral-antineutralon mixing term $H_{n\bar{n}}$ (Eq.~\eqref{mixing}).
Below, we will find the scaling of the length scales associated with $H_{\rm{ch-eq}}$ and $H_{n\bar{n}}$ using the RG method. 
$\ell_0$ acts as the ultraviolet length cutoff of the RG analysis. 

We first consider $\ell_{\rm{ch-eq}, 0}$, characterizing the elastic scattering between the outer-most mode and the inner ones. 
$\ell_{\rm{ch-eq}, 0}$ is determined by the following RG equation, 
\begin{align} \label{Supple:RG1}
\frac{d\tilde{W}_{\rm{ch-eq}}}{d \ln L} = \tilde{W}_{\rm{ch-eq}} (3 - 2 \Delta_{\rm{ch-eq}}).
\end{align}
Here $\tilde{W}_{\rm{ch-eq}} \equiv W_{\rm{ch-eq}} \ell_0^3 / v_n^2 $ (cf. see the inline equation above Eq.~\eqref{effectiveaction} for $W_{\rm{ch-eq}}$) is the dimensionless disorder strength, and $\Delta_{\rm{ch-eq}}$ is the scaling dimension of the $H_{\rm{ch-eq}}$ (Eq.~\eqref{qptunnelingeq}) and is equal to $2/3$ at the fixed point. 
Assuming trivial density of states factors, $\tilde{W}_{\rm{ch-eq}}$ can be thought of as the running dimensionless resistance. 
We now consider the voltage $V$ to be sufficiently small, such that the renormalized 
$\tilde{W}_{\rm{ch-eq}}$ reaches one for $L = \ell_{\rm{ch-eq},0} < L_V$ ($V < V^{*}$ in Fig.~\ref{lengthscales}).
The perturbative RG of Eq.~\eqref{Supple:RG1} breaks down when $\tilde{W}_{\rm{ch-eq}}$ becomes unity and the corresponding length scale (at which $\tilde{W}_{\rm{ch-eq}}$ renormalizes to unity)  is $\ell_{\rm{ch-eq}, 0} = \ell_{0}/\left (\tilde{W}_{\rm{ch-eq}}^{0} \right )^{3/5}$. Here $\tilde{W}_{\rm{ch-eq}}^{0}$ is the bare dimensionless disorder strength.
Following the same procedure for the $H_{n\bar{n}}$ (Eq.~\eqref{mixing}), we also obtain $\ell_{n\bar{n}, 0} = \ell_{0}/\left (\tilde{W}_{n \bar{n}}^{0} \right )^{3/7}$. Both $\ell_{n\bar{n}, 0}$ and $\ell_{\rm{ch-eq}, 0}$ are elastic scattering lengths and do not depend on energy cutoffs (externally applied voltage $V$) of the system. 

As $L_{\textrm{arm}}$ exceeds $\ell_{n\bar{n}, 0}$ and $\ell_{\rm{ch-eq}, 0}$, the system is further renormalized away from the WMG fixed point, ultimately arriving at the low-energy fixed point. In the vicinity of the latter fixed point, counter-propagating neutral modes localize each other, leaving a charge mode and a neutral mode~\cite{Wang2013}. The inelastic length scales (cf. the red curve at $V < V^{*}$ in Fig.~\ref{lengthscales}) near the low-energy fixed point can exceed  the elastic charge equilibration length  $\ell_{\rm{ch-eq}, 0}$ parametrically  at sufficiently small voltages~\cite{Kane1994, Protopopov2017}. By continuity, there exists a regime ($\ell_{\rm{ch-eq, 0}}<L_{\textrm{arm}}<\ell_{\rm{ch-eq}}$) still governed by the WMG fixed point  where our analysis is mostly performed.

When the voltage, on the other hand, is sufficiently large such that $\ell_{\rm{ch - eq}, 0}$ is larger than $L_{V}$ ($V \gg V^{*}$ in Fig.~\eqref{lengthscales}), 
Eq.~\eqref{qptunnelingeq} with Eq.~\eqref{Supple:RG1} yields $\ell_{\rm{ch - eq}}$.
Eq.~\eqref{Supple:RG1} stops to be valid at 
the infrared cutoff $L_V$, leading to 
\begin{align} \label{supple:RG_cheq1}
\frac{\tilde{W}_{\rm{ch-eq}} (L = L_V)}{\tilde{W}_{\rm{ch-eq}}^{0}} = \left (\frac{L_V}{\ell_0} \right)^{5/3}.
\end{align}
Beyond the scale $L_V$, the renormalization of the resistance, hence of $\tilde{W}_{\rm{ch-eq}}$, continues classically (i.e., $\tilde{W}_{\rm{ch-eq}}$ grows linearly as $L$ increases)~\cite{Protopopov2017}, and 
in turn breaks down when $\tilde{W}_{\rm{ch-eq}}$ becomes unity, leading to 
\begin{align} \label{supple:RG_cheq2}
\frac{\tilde{W}_{\rm{ch-eq}} (L = L_V)}{L_V} = \frac{\tilde{W}_{\rm{ch-eq}} (L = \ell_{\rm{ch-eq}})}{\ell_{\rm{ch-eq}}} \simeq \frac{1}{\ell_{\rm{ch-eq}}}. 
\end{align}
Here $\ell_{\rm{ch-eq}}$ is defined as the length at which $\tilde{W}_{\rm{ch-eq}}$ becomes unity. Employing Eqs.~\eqref{supple:RG_cheq1} and \eqref{supple:RG_cheq2}, we obtain $\ell_{\rm{ch-eq}} = \ell_0 (\ell_0 / L_V)^{2/3}/\tilde{W}_{\rm{ch-eq}}^{0} \propto V^{2/3}$ (depicted with the red curve at $V \gg V^{*}$ in Fig.~\ref{lengthscales}). Following the same procedure for the $H_{n\bar{n}}$, we also obtain $\ell_{n \bar{n}} = \ell_0 (\ell_0 / L_V)^{4/3} / \tilde{W}_{n \bar{n}}^{0} \propto V^{4/3}$.

\section{The $Z_3$ symmetry and the two-terminal conductance} \label{appen_z3sym}
In this section, we show self-consistently that the two-terminal conductance is zero if (i) $Z_3$ is present (model (A)) and (ii)
all neutral excitations eventually decay ($\ell_{\rm{ch-eq, 0}} \ll L_{\rm{arm}}$).
\begin{figure} 
\includegraphics[width=0.8\columnwidth]{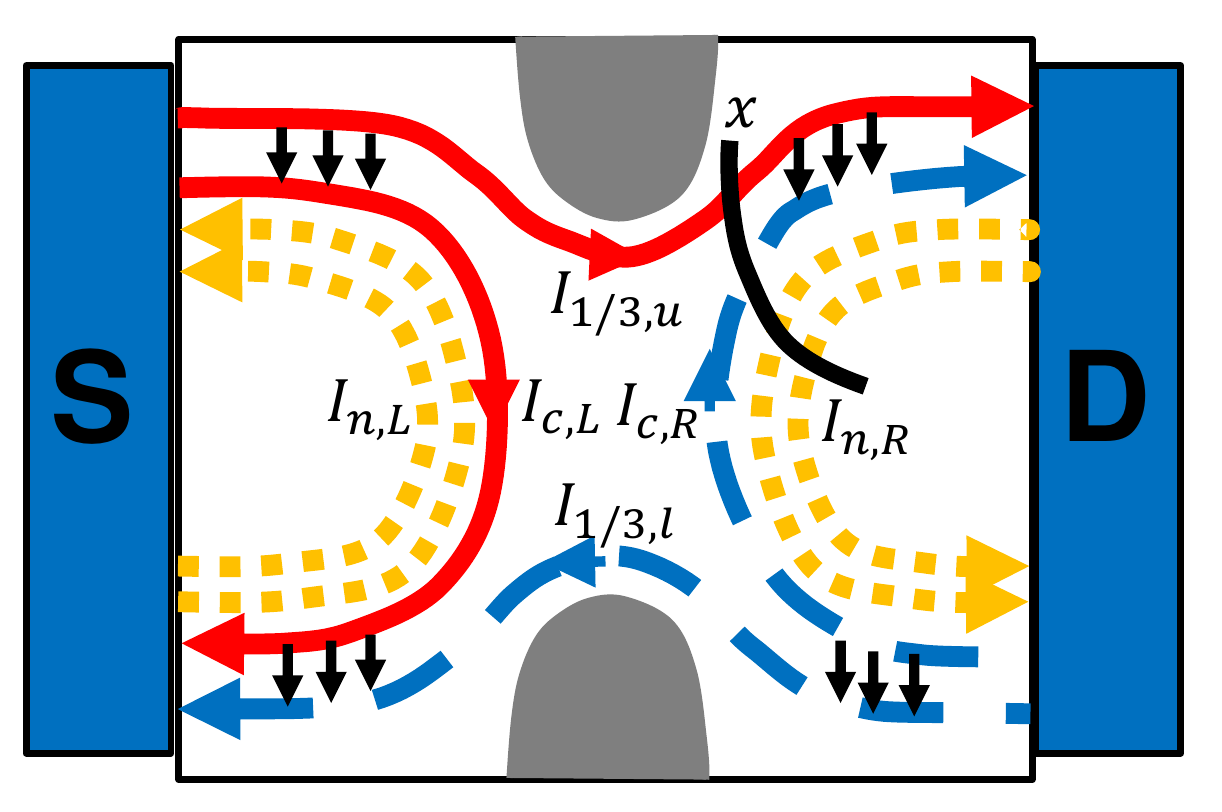} 
\caption{ (color online). A two-terminal setup with a quantum point contact (QPC). Source S is biased by $V_0$. The two-terminal conductance is measured between S and D; the d.c. noise is measured at drain D. Black arrows denote the direction of tunneling charge
currents at each corner.
}\label{T3Sym}
\end{figure}
The quasi-particle tunneling probability between the outer-most mode and the inner modes in each corner is defined as
$P_{u, R}$, $P_{u, L}$, $P_{l, R}$, and $P_{l, L}$, respectively.  
All the electrical currents displayed in Fig.~\ref{T3Sym} are determined by the following rate equations
\begin{align}
I_{1/3, u} &= \frac{e^2 V_0}{3h} + \frac{e}{3} I_{n, L} P_{u, L}, \nonumber\\
I_{1/3, l} &= \frac{e}{3} I_{n, R} P_{l, R},\nonumber \\ 
I_{c, L} &= \frac{e^2 V_0}{3h} - \frac{e}{3} I_{n, L} P_{u, L},\nonumber \\ 
I_{c, R} &= -\frac{e}{3} I_{n, R} P_{l, R}, \nonumber \\ 
\frac{e}{3} I_{n, L} &= \frac{1}{2} P_{l, L} (I_{1/3, l} - I_{c, L}),\nonumber \\ 
\frac{e}{3} I_{n, R}& = \frac{1}{2} P_{u, R}  (I_{1/3, u} - I_{c, R}).
\end{align}
The conductance measured at D is calculated as 
\begin{align} \label{conductance_Z3}
G_{\rm{D}} &\equiv  \frac{ I_{1/3, u}-I_{1/3, l} }{V_0}= \frac{I_{c, R} + I_{1/3, u}}{V_0} \nonumber \\ 
&= \frac{2e^2}{3h} \frac{ (1- P_{l, L}P_{u, L}) (1- P_{l, R}P_{u, R})}{2 - P_{l, L} P_{u, L} - P_{l, R} P_{u, R}}.
\end{align}
When $P = P_{l, L} = P_{l, R} = P_{u, L} = P_{u, R} = 1$, i.e., all the neutral excitations eventually decay ($\ell_{\rm{ch-eq, 0}} \ll L_{\rm{arm}}$), the conductance can be seen to be zero up to an exponential correction of $\sim \exp(-L_{\rm{arm}}/ \ell_{\rm{ch-eq, 0}})$ as $1-P$ goes as $1-P \sim \exp(-L_{\rm{arm}}/ \ell_{\rm{ch-eq, 0}})$. 
When all the tunneling probabilities, on the other hand, go to zero, the conductance becomes $e^2 / (3h)$.

\section{Non-equilibrium noise} \label{appen_noise}

In this section, the noise of neutral currents is shown to be converted into the noise of charge currents by Eq.~\eqref{noiserelation} when all neutral exciations decay ($\ell_{\rm{ch-eq, 0}} \ll L_{\rm{arm}}$). All the calculations in this section are performed employing model (B). 

The densities of neutral modes $n_1$ and $n_2$ are defined as $\rho_{n_1} (x)= \partial_x \phi_{n_1} / ( \sqrt{2} \pi)$ and $\rho_{n_2} = 3\partial_x \phi_{n_2} / (\sqrt{2} \pi)$
such that creation of an {\it up} neutralon changes the density of neutral mode $n_1$ and $n_2$ by a delta function contribution since 
\begin{align} \label{neutraldensity1}
\psi_{u} (x') \Big (\frac{\partial_x \phi_{n_1} (x)}{\sqrt{2} \pi} \Big ) \psi_{u}^{\dagger} (x') 
&=\frac{\partial_x \phi_{n_1} (x)}{\sqrt{2} \pi} + \delta (x - x'), \nonumber \\ \psi_{u} (x')  \Big (\frac{3\partial_x \phi_{n_2} (x)}{\sqrt{2} \pi} \Big )  \psi_{u}^{\dagger} (x') &=\frac{3\partial_x  \phi_{n_2} (x)}{\sqrt{2} \pi} + \delta (x - x').
\end{align} 
Then, we define decay neutral current operators $I_{\textrm{dec}, n_1}$ and $I_{\textrm{dec}, n_2}$ in the decay region of neutralons 
from the equations of motion of neutral number operators 
$N_{n_1} = \int dx \rho_{n_1} (x) $ and $N_{n_2} = \int dx \rho_{n_2} (x) $ as
\begin{align} \label{neutralcurrentoperator2}
I_{\textrm{dec}, n_1} & = 
 \frac{dN_{n_1}}{dt}=\frac{i}{\hbar}[H_{\textrm{eq}}, N_{n_1}] \nonumber \\ 
 &=- \frac{i }{h a}\sum_{\epsilon = \pm} \int dx  \epsilon \left [ T_{u} (x) - T_{d} (x) \right ]^{\epsilon}, \nonumber \\ 
I_{\textrm{dec}, n_2} & = 
 \frac{dN_{n_2}}{dt}=\frac{i}{\hbar}[H_{\textrm{eq}}, N_{n_2}] \nonumber \\ 
 &=-\frac{i }{ha }\sum_{\epsilon = \pm} \int dx \epsilon \left [ T_{u}(x)  + T_{d}(x)  - 2 T_{s}(x) \right ]^{\epsilon},
\end{align}
where the tunneling operator $T_{j = u, d, s}$ of each neutralon is defined as
\begin{align}
T_{j } (x) = t_{j} ( x) \psi_{j}^{\dagger} (x)  e^{- i \phi_c(x) /\sqrt{3}} e^{ i\phi_{1/3} (x)},
\end{align}
and a convenient notation $   [AB]^{+} = AB$ ($[AB]^{-} = B^{\dagger} A^{\dagger}$) is used. Similarly, a charge tunneling current is defined as
\begin{align} \label{currentoperator2}
I_{\textrm{tun}} & =\frac{ie}{\hbar}\left [H_{\textrm{eq}},  \frac{1}{2 \pi }\int dx \partial_x \phi_{1/3} \right]
\nonumber \\ 
&=\frac{i e}{3 h a}\sum_{\epsilon = \pm} \sum_{j = u, d, s}  \int dx \epsilon  \left   [ T_u (x) + T_d (x) + T_s (x) \right ]^{\epsilon}.
\end{align}
The decay neutral currents and incoming neutral currents are zero, $\langle I_{\textrm{dec}, n_1} \rangle = \langle I_{\textrm{dec}, n_2} \rangle = 0 $ and $\langle I_{n_1} \rangle = \langle I_{n_2} \rangle =0 $ (for their definitions, see in-line equations just below Eq.~\eqref{correlators}); $\langle I_{\textrm{dec}, n_1} \rangle = \langle I_{\textrm{dec}, n_2} \rangle = 0$ can be derived using the fact that 
the Green's function of neutralons (Eq.~\eqref{noneqGreenmodelB}) is real for the model (B). $\langle I_{n_1} \rangle = \langle I_{n_2} \rangle =0 $ can be derived (i) taking the first derivative of Eqs.~\eqref{generatingFun1} and \eqref{generatingFun2} by $\lambda_1$ and $\lambda_2$ and (ii) sending $\lambda_1$ and $\lambda_2$ to zero.
Furthermore, the electrical tunneling current is zero as seen in Eq.~\eqref{noneqGreenmodelB}. 
Under the assumption that all neutral excitations eventually
decay in the decay region, the noise ($S_{n_1}$ and $S_{n_2}$) of incoming neutral currents are identical to the noise of the decay neutral currents, 
\begin{align} \label{neutralnoises}
S_{n_1} &= S_{\rm{dec}, n_1} \equiv 2 \int_{-\infty}^{\infty} dt \left \langle I_{\textrm{dec}, n_1} (t) I_{\textrm{dec}, n_1} (0)\right \rangle \nonumber \\
S_{n_2} &= S_{\rm{dec}, n_2} \equiv 2 \int_{-\infty}^{\infty} dt \left \langle I_{\textrm{dec}, n_2} (t) I_{\textrm{dec}, n_2} (0)\right \rangle.
\end{align}
respectively. Using Eqs.~\eqref{neutralcurrentoperator2}-\eqref{neutralnoises}, the noise ($S_{\textrm{tun}}$) of the electical tunneling current can be decomposed into the noise of the neutral decay currents as
\begin{widetext}
\begin{align}
S_{\textrm{tun}} \equiv & 2 \int dt \left \langle I_{\textrm{tun}}(t)  I_{\textrm{tun}}(0) \right \rangle  \nonumber \\ 
\simeq &\frac{e^2}{9 \hbar^2 \pi a } \sum_{\epsilon = \pm} \int dt \int dx \int dx' \left (
\left \langle \left [T_s (x, t) \right]^{\epsilon} \left [T_{s}^{\dagger} (x', 0) \right]^{\epsilon}\right \rangle  
+ \left \langle \left [T_u (x, t) \right]^{\epsilon} \left [T_{u}^{\dagger} (x', 0) \right]^{\epsilon}\right \rangle  
+\left \langle \left [T_d (x, t) \right]^{\epsilon} \left [T_{d}^{\dagger} (x', 0) \right]^{\epsilon}\right \rangle   
\right ) \nonumber \\ 
=& \frac{e^2}{9 \hbar^2 \pi a }  \sum_{\epsilon = \pm} \int dt \int dx \int dx' 
\bigg [ \frac{3}{4} \left (
\left \langle \left [T_u (x, t) \right]^{\epsilon} \left [T_{u}^{\dagger} (x', 0) \right]^{\epsilon}\right \rangle  
+\left \langle \left [T_d (x, t) \right]^{\epsilon} \left [T_{d}^{\dagger} (x', 0) \right]^{\epsilon}\right \rangle   
\right ) \nonumber \\
 & +  \frac{1}{4} \left (
\left \langle \left [T_u (x, t) \right]^{\epsilon} \left [T_{u}^{\dagger} (x', 0) \right]^{\epsilon}\right \rangle  
+\left \langle \left [T_d (x, t) \right]^{\epsilon} \left [T_{d}^{\dagger} (x', 0) \right]^{\epsilon}\right \rangle  
+ 4   \left \langle \left [T_s (x, t) \right]^{\epsilon} \left [T_{s}^{\dagger} (x', 0) \right]^{\epsilon}\right \rangle  
\right )  \bigg ] \nonumber \\  = &\frac{e^2}{36} (3S_{\textrm{dec}, n_1} +S_{\textrm{dec}, n_2} ) = \frac{e^2}{36} (3S_{n_1} +S_{n_2} ) .
\end{align}
\end{widetext}
This charge noise of the electrical tunneling current is measured at drain D (see Fig.~2).
The charge noise in each of the upper and lower edges contributes to the zero-frequency noise $S_{\rm{D}}$ as
\begin{align} \label{appendix_noiserelation}
 S_{\textrm{D}}= \frac{e^{2}}{36} (3 S_{n_1, \textrm{u}} +  S_{n_2, \textrm{u}}
+ 3 S_{n_1, \textrm{l}} + S_{n_2, \textrm{l}}) = \frac{2e^3 V}{9h},
\end{align}
arriving at Eq.~\eqref{noiserelation}.

\end{document}